\documentclass[12pt]{article}
\pdfoutput=1
 \usepackage{latexsym, graphicx} 

\newcommand{\arXiv}[1]{\href{http://www.arXiv.org/abs/#1}{#1}}
\usepackage[colorlinks=true, linkcolor=blue, bookmarks=true]{hyperref}

\makeatletter
\renewcommand\section{\@startsection {section}{1}{\z@}%
                                   {-3.5ex \@plus -1ex \@minus -.2ex}
                                   {2.3ex \@plus.2ex}%
                                   {\normalfont\large\bfseries}}
\renewcommand\subsection{\@startsection{subsection}{2}{\z@}%
                                     {-3.25ex\@plus -1ex \@minus -.2ex}%
                                     {1.5ex \@plus .2ex}%
                                     {\normalfont\bfseries}}
\makeatother
\newcommand{\beq}{\begin{equation}}
\newcommand{\eeq}{\end{equation}}
\newcommand{\ber}{\begin{array}}
\newcommand{\eer}{\end{array}}
\newcommand{\D}{{\cal D}}

\newcommand{\dtwo}{d^{\hspace{1pt}2}\hspace{-1pt}}
\newcommand{\del}{\partial}

\newcommand{\sssty}{\scriptscriptstyle}
\newcommand{\ssty}{\scriptstyle}
\newcommand{\dsty}{\displaystyle}
\newcommand{\s}{\sigma}
\newcommand{\te}{\theta}

\newcommand{\de}{\delta}
\newcommand{\ds}{\dtwo\sigma}
\newcommand{\cnst}{\mbox{const}}

\newcommand{\eps}{\varepsilon}

\newcommand{\ena}{\end{eqnarray}}
\newcommand{\beqa}{\begin{eqnarray}}
\newcommand{\eeqa}{\end{eqnarray}}
\newcommand{\bea}{\begin{eqnarray}}
\newcommand{\eea}{\end{eqnarray}}

\newcommand{\be}{\begin{equation}}
\newcommand{\ee}{\end{equation}}
\newcommand{\uG}{\bar G}

\newcommand{\uX}{\underline{X}}

\newcommand{\bx}{\Box}

\begin{document}
\begin{titlepage}
\begin{flushright}
\phantom{arXiv:yymm.nnnn}
\end{flushright}
\vfill
\begin{center}
{\Large\bf Strings in compact cosmological spaces}    \\
\vskip 15mm
{\large Ben Craps$^a$, Oleg Evnin$^b$ and Anatoly Konechny$^c$}
\vskip 7mm
{\em $^a$ Theoretische Natuurkunde, Vrije Universiteit Brussel and\\
The International Solvay Institutes\\ Pleinlaan 2, B-1050 Brussels, Belgium}
\vskip 3mm
{\em $^b$ Department of Physics, Faculty of Science, Chulalongkorn University,\\
Thanon Phayathai, Pathumwan, Bangkok 10330, Thailand}
\vskip 3mm
{\em $^c$ Department of Mathematics, Heriot-Watt University and\\
Maxwell Institute for Mathematical Sciences, Edinburgh EH14 4AS, UK}

\vskip 3mm
{\small\noindent  {\tt Ben.Craps@vub.ac.be, oleg.evnin@gmail.com, A.Konechny@hw.ac.uk}}
\vskip 10mm
\end{center}
\vfill

\begin{center}
{\bf ABSTRACT}\vspace{3mm}
\end{center}

We confront the problem of giving a fundamental definition to perturbative string theory in spacetimes with totally compact space (taken to be a torus for simplicity, though the nature of the problem is very general) and non-compact time. Due to backreaction induced by the presence of even a single string quantum, the  usual formulation of perturbative string theory in a fixed classical background is infrared-divergent at all  subleading orders in the string coupling, and needs to be amended. The problem can be seen as a closed string analogue of D0-brane recoil under an impact by closed strings (a situation displaying extremely similar infrared divergences). Inspired by the collective coordinate treatment of the D0-brane recoil, whereby the translational modes of the D0-brane are introduced as explicit dynamical variables in the path integral, we construct a similar formalism for the case of string-induced gravitational backreaction, in which the spatially uniform modes of the background fields on the compact space are quantized explicitly. The formalism can equally well be seen as an ultraviolet completion of a minisuperspace quantum cosmology with string degrees of freedom. We consider the amplitudes for the universe to have two cross-sections with specified spatial properties and string contents, and show (at the first non-trivial order) that they are finite within our formalism.

\vfill

\end{titlepage}

\section{Introduction}

One is used to formulating perturbative string theory in a fixed classical background. The absence of Weyl anomalies on the worldsheet then implies that the equations of motion for the background space-time fields are satisfied (and, to the lowest order in the derivative expansion, these equations coincide with supergravity equations of motion). The objective of this article is to investigate important cases when such separation into a fixed classical background and stringy excitations is fundamentally invalid.

In a situation when even one string exerts substantial backreaction on the classical background, one cannot expect the usual formulation of string perturbation theory to be viable. Indeed, a single string is a fundamentally quantum object, without well-defined coordinates and velocities, and its backreaction does not amount to a classical deformation of the background. In cases when this backreaction is substantial, the very notion of classical background will be compromised.

A single string excitation will exert significant backreaction on those backgrounds that are ``small'' in some sense. The ``smallness'' may mean a small number of non-compact or world-volume dimensions, as we shall immediately see in the following examples.

Perhaps the simplest case when the need for a radical modification of the usual string perturbation theory becomes apparent is closed string scattering in the background of a D0-brane. 
As a result of the impact by the incident closed strings, the D0-brane starts to move with a constant velocity and ultimately deviates infinitely far from the original static D0-brane configuration used in formulating string perturbation theory. Furthermore, since closed strings are inherently quantum and cannot have a well-defined spatial position, the moment of impact is likewise not well-defined and the D0-brane (hit at a moment bearing quantum uncertainty) will not move on a fixed classical trajectory. (In situations when the D0-brane is hit by a well-localized closed string wave-packet, its subsequent trajectory may be nearly classical, but this is never exact, due to the uncertainty principle.)

One might think that the above concerns are lyrical in nature, and the theory will automatically take care of the backreaction effects (such attitude may, in particular, be fueled by the common catch-phrase that ``string excitations describe deformations of the background''). One could therefore try to proceed na\"\i vely with a traditional formulation of string perturbation theory for the case of recoiling D0-branes, only to discover that all sub-leading orders of the string coupling expansion are infrared-divergent, and the theory is not useable as is. Not surprisingly, the divergences bear close resemblance to the more familiar field-theoretical soliton recoil problem \cite{rajaraman,christ-lee}.

When infrared divergences are encountered in string perturbation theory, one usually looks for an implementation of the Fischler-Susskind mechanism \cite{fs}, i.e., finding a small deformation of the background that introduces small additional divergences on lower-genus worldsheets that cancel the pre-existing infrared divergences on higher-genus worldsheets. In the familiar cases (e.g., bosonic string divergences induced by tadpoles), the said background deformation simply amounts to replacing the given classical space-time configuration with a slightly different one. For the situation we are dealing with, one cannot expect this strategy to work, because the recoiling D0-brane does not move along a fixed classical trajectory. Hence, the implementation of the Fischler-Susskind mechanism has to be altered accordingly.

Motivated by the collective coordinate treatment of the soliton recoil problem  \cite{rajaraman,christ-lee}, one may try to introduce the center-of-mass coordinates of the D0-brane as explicit dynamical variables and perform functional integration over them in addition to the usual functional integration over the string worldsheets (attached to the D0-brane worldline). The presence of curved D0-brane worldlines in the functional integral introduces additional ultraviolet divergences in the worldsheet path integral that should cancel the infrared divergences associated with recoil. It has been explicitly demonstrated in \cite{d0,thesis} that this cancellation indeed happens at the first non-trivial order in the string coupling, in accordance with the intuition that taking a proper account of the recoil effect cures the theory of infrared divergences.

We shall now turn to our second example of backreaction that invalidates conventional string perturbation theory, which will be the main focus of this paper. One can consider strings propagating in a space-time with some compactified spatial dimensions (for simplicity, we shall concentrate on compactifications of torus topology, though the issues are completely general). If the number of compactified spatial dimensions is not too large, a finite number of strings always induce a zero average energy density, and deform the background only locally, close to the region where they propagate. Far away from the string scattering experiment, the space-time always stays as is, and the space-time background used in formulating string perturbation theory remains valid. The situation is completely different if {\it all} spatial dimensions are compactified. In this case, even a single string induces a non-zero average energy density, and the space-time will obviously undergo cosmological evolution, driving it far away from the original classical configuration assumed in the conventional perturbative expansion of string theory.\footnote{Peculiar strong backreaction effects accompanied by divergences also exist for exactly one non-compact spatial dimension. For their analysis in the context of scattering off stretched D1-branes, see \cite{localrecoil}.} Just as for the recoil case, since the source of backreaction is inherently quantum, its effect will not amount to any fixed classical deformation of the background space-time.

In close analogy to the recoil case, the backreaction effect from a finite number of quantum strings does not merely present a new qualitative twist to the usual string physics. Rather, it introduces infrared divergences in string perturbation theory, thereby invalidating all subleading orders in string coupling. Needless to say, the case of gravitational backreaction is physically more interesting than the case of recoil, due to its cosmological implications. Indeed, a number of intriguing conjectures (see, e.g., \cite{bv}, and \cite{sgrev1,sgrev2} for reviews) have been made in relation to the evolution of compact spaces filled with string gases. Our present objective is to give a microscopic definition of perturbative string theory in such a setting (given that the conventional formulation breaks down at all subleading orders in the string coupling due to infrared divergences).

Inspired by the success of the collective coordinate treatment of the recoiling D0-brane, one may try to repeat the procedure for the gravitational case (some preliminary considerations in this regard have been given in \cite{qbackreact}). Inspecting the infrared divergences responsible for the breakdown of the conventional string perturbation theory, one discovers that they are induced by the zeroth (uniform) Kaluza-Klein modes of the massless fields present in string theory (metric, dilaton, B-field; we are concentrating on the case of bosonic string theory for simplicity, though the key issues will be equally relevant in any superstring theory). Introducing an explicit path integral over the said modes, one ends up with bosonic strings moving in a minisuperspace universe of topology $R\times T^{25}$. The formalism we are dealing with can equally well be seen as an ultraviolet completion of a minisuperspace cosmological model with stringy degrees of freedom. The main purpose of our present investigation is to show that the infrared divergences indicative of gravitational backreaction are cured in the framework of our present treatment.

Interactions of strings in a compact cosmological space have been considered from a phenomenological perspective in \cite{intcom} (which can also be consulted for a large list of references to earlier publications). There, simplified models were devised to give effective description of winding string annihilation (crucial for the Brandenberger-Vafa scenario \cite{bv}). Our aim is rather to develop a systematic worldsheet treatment (which may, in particular, shed light on the validity of the phenomenological description in \cite{intcom}).

The infrared divergences from small worldsheet handles in compact target space-times (different from, but related to compact target spaces we are handling here) were also considered in \cite{Friedan_IR}. In that paper, the divergence was treated by introducing new {\it worldsheet} degrees of freedom, precisely designed to cancel the divergences. The approach is quite general and might apply in our setting as well.
The treatment of \cite{Friedan_IR} preserves the usual worldsheet factorization properties that underlie the standard unitarity arguments for perturbative string theory.  Our approach, on the other hand, has a
transparent space-time interpretation, as the  degrees of freedom that receive the special treatment are simply the long wavelength fluctuations of the target space-time  background fields (rather than new worldsheet degrees of freedom specifically designed to cancel the divergences). Also as the D0-brane recoil problem clearly 
shows, the unitarity at stake is no longer the  unitarity of the string perturbation series alone but rather includes non-perturbative degrees of freedom. 

The paper is organized as follows: we first review the structure of the infrared divergences invalidating conventional string perturbation theory in section 2 (both for D0-brane recoil and for gravitational backreaction in a compact cosmological space) and the collective coordinate treatment of the D0-brane recoil problem in section 3. In section 4, we present the collective coordinate formalism for the gravitational case and give a crude qualitative sketch of the overall structure of potentially divergent terms. In section 5, we provide a more detailed and technical treatment of the transition amplitudes in our string cosmology formalism, establishing finiteness in greater detail.


\section{Analogy between gravitational backreaction in compact spaces and soliton recoil}

We shall now review the infrared divergences arising in the conventional formulation of string theory, and draw close parallels between the cases of D0-brane recoil and gravitational backreaction in a totally compact space. It is convenient to start with the recoil problem, which is more straightforward in a number of ways.

\begin{figure}[t]
\centering
\includegraphics[width=10cm]{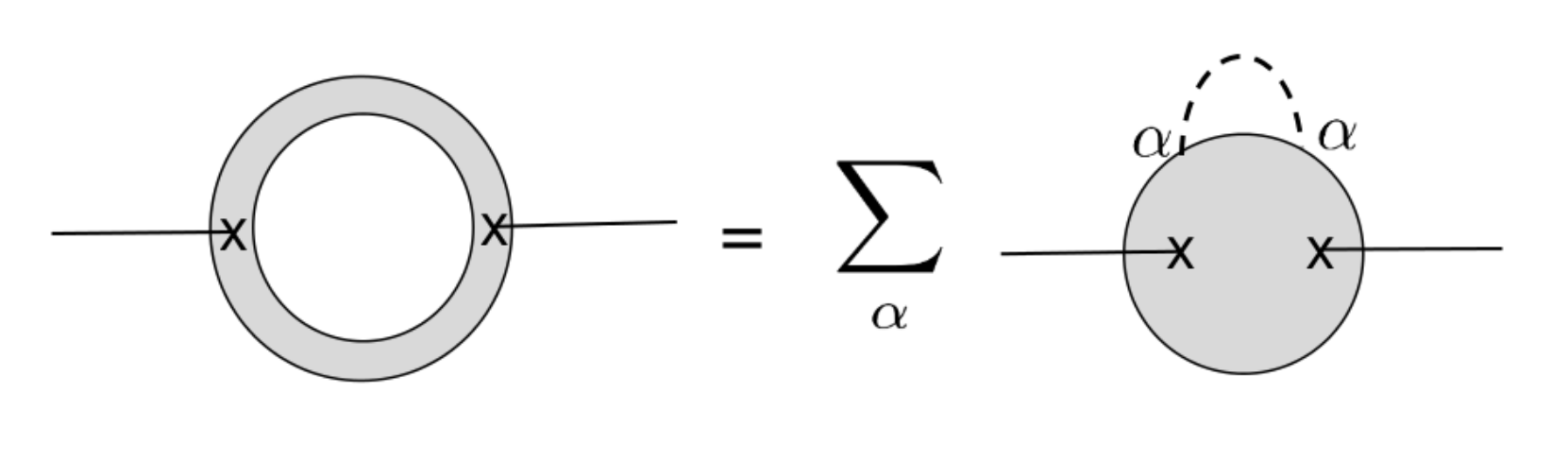}
\caption{The plumbing fixture constuction applied to an annulus worldsheet.}
\label{plmb_open}
\end{figure}

As briefly alluded to in the introduction, one could try to formulate the conventional string perturbation theory in the background of a straight D0-brane worldline, only to discover that the perturbation theory is infrared-divergent in all subleading orders. The first divergence, which we shall now display explicitly, comes at next-to-leading order in the string coupling from annulus worldsheets developing a long, thin strip. 
Divergences from degenerating Riemann surfaces can be analyzed using Polchinski's ``plumbing fixture'' construction \cite{polch, polchinski-fischler-susskind}, which relates the divergences to amplitudes evaluated on  lower genus Riemann surfaces. 
In particular, the annulus amplitude with an insertion of closed string vertex operators $V^{(1)},\cdots,V^{(n)}$ can be expressed in terms of  disk amplitudes with additional operator insertions at the boundary:
\beq
\left\langle V^{(1)}\cdots V^{(n)}\right>_{annulus}=\sum\limits_\alpha \int \frac{dq}q\, q^{h_\alpha-1} \int d\te d\te' \left\langle V_\alpha(\te)V_\alpha(\te')V^{(1)}\cdots V^{(n)}\right>_{disk}\, . 
\label{plumb}
\eeq
Here the summation goes over a complete set of local operators $V_\alpha(\te)$ with conformal weights $h_\alpha$, and $q$ is the gluing parameter  related to the annular modulus. ($\te$ parametrizes the boundary of the disk.) 
The divergence in the integral over $q$ that comes from the region $q\approx0$ (i.e., from an annulus developing a thin strip) is dominated by the terms with the smallest possible $h_\alpha$. 
A schematic representation of the manipulation with worldsheets leading to (\ref{plumb}) is given in Fig.~\ref{plmb_open}.

Neglecting the tachyon divergence, which is absent for superstring models, we consider the following operators with conformal weights $h=1+\alpha'\omega^2$:
$V^i(\te)= \del_n X^i(\te)\exp\left[i\omega X^0(\te)\right]$. These operators correspond to massless open string states  that represent translations of the D0-brane in the $i$'th Dirichlet direction.  
For small values of $q$, which is the region we are interested in, only small values of $\omega$  contribute to the integral. Hence, the annular divergence takes the following form:
\bea
\left\langle V^{(1)}\cdots V^{(n)}\right>_{annulus}^{(div)}\sim\int\limits_0^1 dq\int\limits_{-\infty}^\infty  d\omega\, q^{-1+\alpha'\omega^2}\int d\te d\te' \left\langle V^i(\te,\omega)V^i(\te',\omega)V^{(1)}\cdots V^{(n)}\right>_{disk}\nonumber\\
\phantom{.}\hspace{-3cm}\sim P^2\left\langle V^{(1)}\cdots V^{(n)}\right>_{disk}\int\limits_0^1 dq\int\limits_{-\infty}^\infty d\omega\, q^{-1+\alpha'\omega^2}\sim P^2\left\langle V^{(1)}\cdots V^{(n)}\right>_{disk}\int\limits_0^1 {dq\over q\,(-\log q)^{1/2}}\ ,
\label{annulus}
\eea
where we have taken into account the fact that the operator $\int \del_n X^i(\te)d\te$ merely shifts the position of the D0-brane.  Inserting this operator into any amplitude amounts to multiplication by the total (Dirichlet) momentum $P$ transferred by the closed strings to the D0-brane during scattering. Introducing a cut-off $\eps$ on the lower bound of the $q$-integral in (\ref{annulus}) reveals a  $\sqrt{|\log\eps|}$ divergence that is characteristic of recoil:
\beq
\left\langle V^{(1)}\cdots V^{(n)}\right>_{annulus}^{(div)}\sim P^2\left<V^{(1)}\cdots V^{(n)}\right>_{disk}\sqrt{|\log\eps|}\, .
\label{annulusD0}
\eeq
The fact that the divergence is proportional to the square of the transferred momentum highlights its relation to recoil (the divergence vanishes when no momentum is transferred). 
The overall normalization can be determined with additional effort (see \cite{d0}). (Interestingly, a related divergence also occurs for a stretched D1-brane, but not for higher-dimensional branes, indicating a peculiar backreaction effect related to the special features of low-dimensional kinematics, see \cite{localrecoil}.)


We now turn to the case of strings in a totally compact cosmological space. Just as for the case of D0-brane recoil, it is important to see how na\"\i ve attempts to formulate string perturbation theory fail.

For example, one can consider strings on a flat space-time with all spatial directions compactified on circles ($R\times T^{25}$ for bosonic string theory). One often hears that massless string states describe deformations of the background (and this intuition is perfectly sound for non-compact Minkowski space backgrounds), however for this case when quantum backreaction occurs, conventional perturbative string theory is plainly divergent and fails to account for the cosmological dynamics of the compact background space.

\begin{figure}[t]
\centering
\includegraphics[width=10cm]{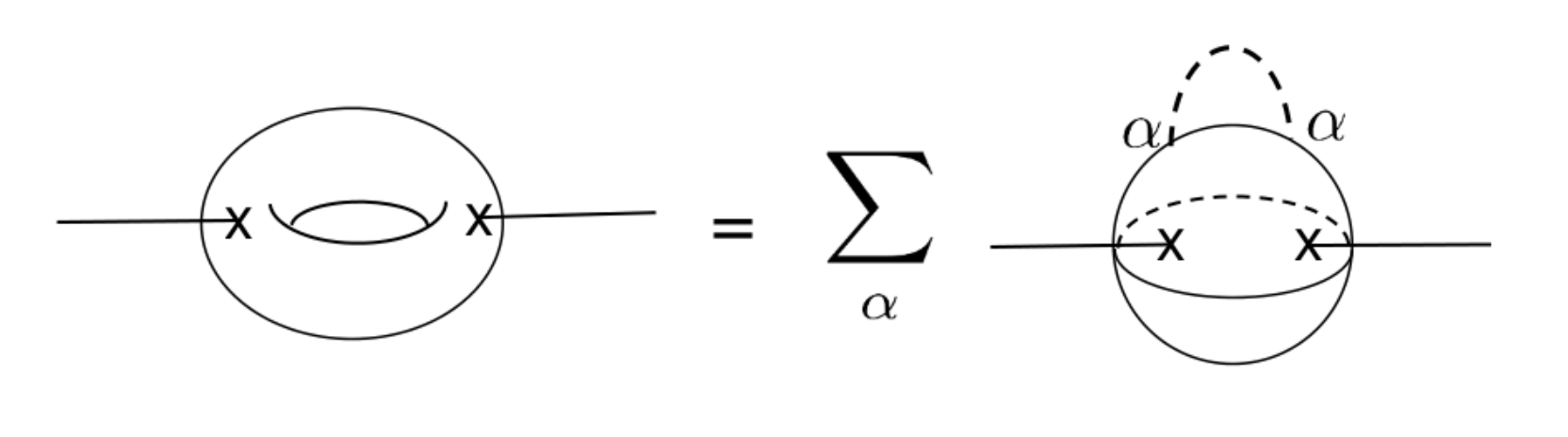}
\caption{The plumbing fixture constuction applied to a torus worldsheet.}
\label{plmb_closed}
\end{figure}

Just as in the D0-brane recoil case, the first divergence comes at next-to-leading order in the string coupling, which corresponds to a torus worldsheet. More specifically, the divergence arises from the limit when the torus develops a long thin handle (described in this case by a single complex modular parameter $q$). As usual, via the plumbing fixture construction \cite{polch, polchinski-fischler-susskind}, the divergence can be expressed through sphere amplitudes with two additional vertex operator insertions corresponding to massless string states carrying zero momentum in all compact directions, multiplied by propagator factors
\beq
\int d\omega\int_{|q|<1}\frac{d^2q}{q\bar q}(q\bar q)^{\alpha'\omega^2/4}\sim\int \frac{d\kappa}{\kappa(-\log\kappa)^{1/2}},
\label{cosmdiv}
\eeq
where $\kappa=(q\bar q)^{1/2}$. Cutting off the lower bound of the $\kappa$-integration at $\kappa=\eps$ reveals, just as it did in the recoil case, a $\sqrt{|\log\eps|}$ divergence. (A schematic representation of the manipulation with worldsheets leading to (\ref{cosmdiv}) is given in Fig.~\ref{plmb_closed}.)

It may appear surprising that such physically distinct processes as soliton recoil and cosmological backreaction appear on the same footing in our present treatment. Nonetheless, it should be apparent that the structure of the infrared divergences in string perturbation theory associated with the two backreaction effects is nearly identical (the two cases under consideration simply being the open and closed string versions of the same mathematical structure).

It is a customary perception that infrared divergences in string theory require a background modification that would introduce extra ultraviolet divergences on lower-genus worldsheets that would cancel the pre-existing infrared ones (this paradigm goes under the name of the Fischler-Susskind mechanism). Our treatment will largely conform to this expectation, except that the `background deformation' will not amount to replacing one classical background with another (intuitive reasons for why that is so have been given in the introduction). Rather, we shall integrate over a class of backgrounds described by a set of collective coordinates, a class rich enough to accommodate the backreaction effects. The criterion by which we judge this procedure is the finiteness properties of the ensuing amplitudes.


\section{A brief review of the worldline formalism and recoil perturbation theory for D0-branes}

\subsection{Worldline path integral}

Having displayed the leading infrared divergence that invalidates the conventional string perturbation theory in the presence of a recoiling D0-brane, we shall now discuss the possible cures. One might have tried to introduce a classical background deformation, i.e., a classical recoiling (curved) D0-brane in the spirit of the usual applications of the Fischler-Susskind mechanism. Apart from being intuitively wrong (because in the physical recoil induced by an impact of quantum string scattering states the resulting trajectory can never be exactly classical), any classical deformation of the D0-brane trajectory will necessarily break time translation symmetry, and the system of a recoiling D0-brane together with incident closed strings clearly respects time translation invariance (and conserves energy).\footnote{In fact, precisely this approach has been tried in \cite{pt}, and mathematically, their treatment did contain some relevant elements that form a close parallel with ours, but it is ultimately unsatisfactory for the reasons we have just described.}

Let us turn for a moment to the problem of soliton recoil (a field theory analogue of the D0-brane recoil). For that case, one has the advantage of having the underlying Lagrangian formulation, from which various perturbative expansions may be derived. It is known 
\cite{rajaraman, christ-lee} that the field theory path-integral in the one-soliton sector can be recast in a form where the center-of-mass coordinate of the soliton appears as an explicit dynamical variable, and the remaining path integral is performed only over field modes orthogonal to the translational mode of the soliton. The perturbative expansion in this setting is manifestly free of infrared divergences (whereas without explicit separation of the translational modes of the soliton, infrared divergences arise in all subleading orders of the perturbative expansion, much in the same manner as for the D0-brane recoil case).

For the case of D0-brane recoil, we are lacking a fundamental underlying theory from which string perturbation theory (or its modifications) can be derived. The best we can do is to try introducing an ad hoc modification of string perturbation theory modelled on the soliton recoil perturbation theory briefly introduced above, and judge it by whether it meets various consistency requirements (such as divergence cancellation). To this end, we shall introduce an explicit path integral over the D0-brane worldline, denoted $f^\mu(t)$, in addition  to the usual path integral over all worldsheets, denoted $X^\mu(\sigma)$, attached to the worldline via Dirichlet boundary conditions. (A formalism of this sort was proposed in \cite{hk}, substantially reorganized and strengthened in \cite{worldline} and finally applied to the recoil problem in the form we are presently reviewing in \cite{d0, thesis}.)

\begin{figure}[t]
\centering
\includegraphics[width=10cm]{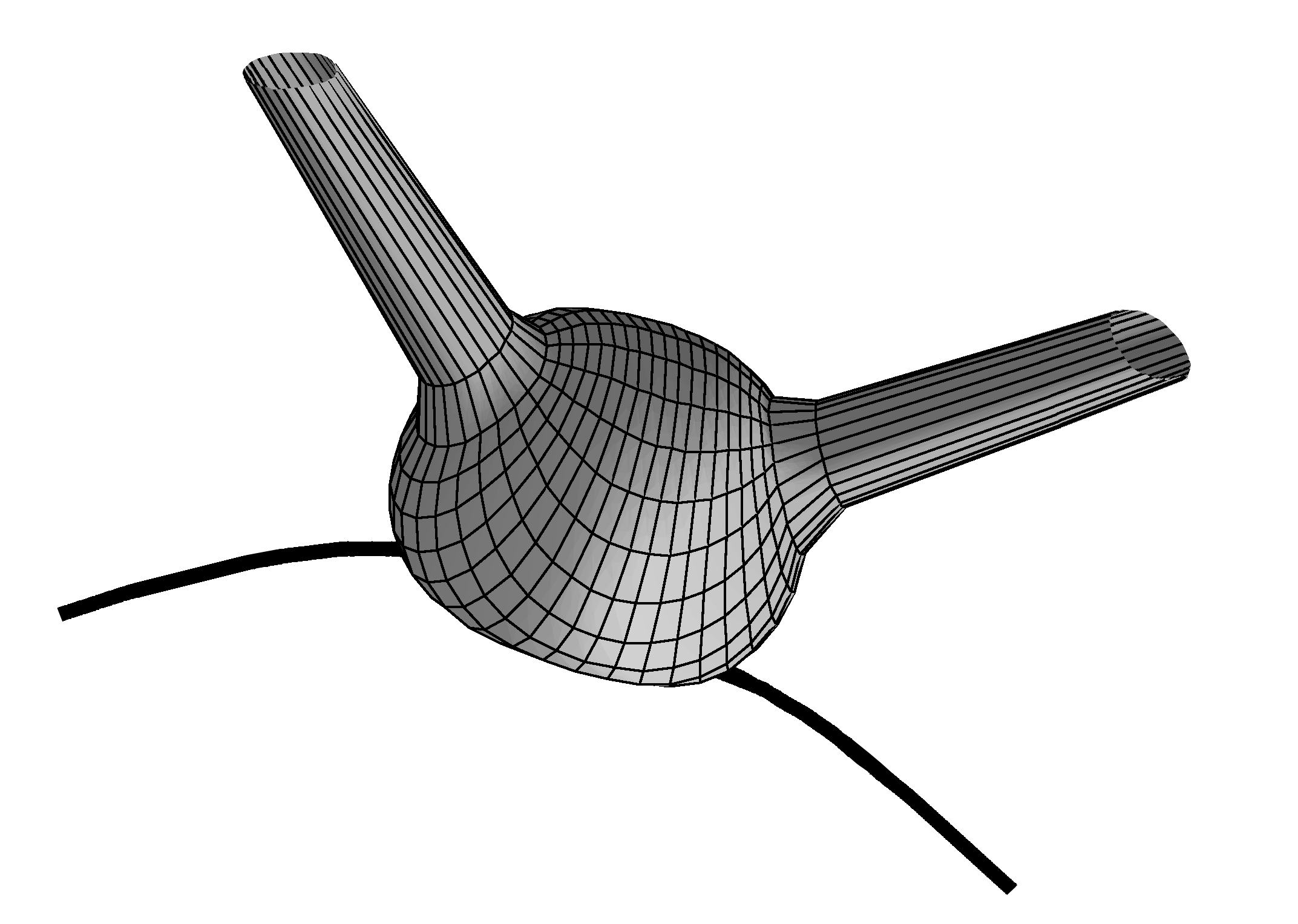}
\caption{A dynamical recoiling D0-brane worldline coupled to a string worldsheet.}
\label{recl}
\end{figure}

The main physical object  we consider is the amplitude for a D0-brane to move from the point $x_1^\mu$ to the point $x_2^\mu$ while absorbing/emitting closed strings with momenta $k_n$:
\beq
\ber{l}
\dsty G(x_1,x_2|\,k_n)=\sum\left(g_s\right)^\chi \int \D f\, \D t\,\D X\,\de\left(X^\mu(\te)-f^\mu(t(\te))\right)e^{-S_{D}(f)-S_{st}(X)} \prod\left\{g_s V^{(n)}(k_n)\right\}\, . 
\eer
\label{master}
\eeq
Here $S_{st}$ is the standard conformal gauge string action
($4\pi\alpha'S_{st}=\int\ds\nabla X_\mu\nabla X^\mu$),
the integration with respect to $f_\mu$ extends over all the inequivalent (i.e., unrelated by diffeomorphisms) curves starting at $x_1$ and ending at $x_2$, $\te$ parameterizes the boundary of the worldsheet, 
and $t(\te)$ describes how this boundary is mapped onto the D0-brane worldline. The summation  runs over all of the topologies of the worldsheets (not necessarily connected, but without any disconnected vacuum parts) and $\chi$ stands for the Euler number. 
 Since we shall be mostly working with worldsheets of disk topology the integration over moduli
of the worldsheet is suppressed. The scattering amplitude can be obtained from the Green's function (\ref{master}) by means of the standard reduction formula:
\beq
\left<p_1|p_2\right>_{k_n}=\lim\limits_{p_1^2,p_2^2\to -M^2} \left(p_1^2+M^2\right)\left(p_2^2+M^2\right)\int dx_1dx_2 e^{ip_1x_1} e^{ip_2x_2}G(x_1,x_2|\,k_1,\cdots,k_m), 
\label{reduct}
\eeq
where $M$ is the D0-brane mass. The choice of the worldline action $S_D$ is subtle (see \cite{d0,thesis,hk,worldline} for some discussion) but for the lowest order consideration it suffices to consider the simplest free-particle action 
(the length of the worldline times the  mass of the D0 brane): $S_D[f(t)]=MT$. An illustration of the kind of configurations entering the above path integrals is given in Fig.~\ref{recl}.

\subsection{Divergence cancellation}

It is important to realize that the path integral (\ref{master}) contains contributions from curved D0-brane worldlines.   Curved  worldlines will lead to ultraviolet divergences in the worldsheet path integral  and to  Weyl anomalies. 
This situation is closely analogous to a more familiar case: string propagation in a curved space-time. Since the D0-brane is heavy (its mass is of the order of $1/g_{s}$)  the typical curvature of its worldline is small. 
Hence the divergences (on the disk, for the purposes of our considerations) will gain an additional power of $g_s$, which will put them in the right order of the string coupling to cancel the annulus divergence (\ref{annulusD0}).

 Intuitively we expect   that making the D0-brane worldlines dynamical automatically cancels the annulus divergence that invalidates conventional string perturbation theory. 
 This  proves to hold true upon evaluation of the path integral (\ref{master}). We refer the reader to \cite{d0,thesis} for detailed derivations. Schematically, one finds the following relation between the divergent part of the next-to-leading order contribution to the disk 
 scattering amplitude (\ref{reduct}), $\left<p_2|p_1\right>^{(1;div)}$, and the (finite) leading order contribution to the same amplitude, $\left<p_2|p_1\right>^{(0)}$:
\beq
\left<p_2|p_1\right>^{(1;div)}\sim\frac{P^2}{M}\left<p_2|p_1\right>^{(0)}\int d\omega\, G_{kick}(\omega)e^{2\alpha'\omega^2\log\de}\sim\sqrt{\alpha'|\log\de|}\,\,\frac{P^2}{M}\left<p_2|p_1\right>^{(0)},
\label{diskdiv}
\eeq
where $G_{kick}(\omega)\sim\frac{1}{(\omega+i0)^2}+\frac{1}{(\omega-i0)^2}$ is the Fourier transform of the Green function of free motion on a line $G_{kick}(t)=|t|/2$, and $\de$ is a small-distance worldsheet cut-off. One can recognize a structure similar to (\ref{annulusD0}) and matching momentum dependences, and upon closer inspection, with an appropriate  identification of minimal distance cut-off on the disk and modulus cut-off on the annulus, $\left<p_2|p_1\right>^{(1;div)}$ exactly cancels the divergent contribution to the same process coming from the annulus, thus successfully implementing a Fischler-Susskind type mechanism at the next-to-leading order in the string coupling.

We note  for the sake of future discussion that (\ref{diskdiv}) manifestly contains contributions of different orders in $\alpha'$ whereas in familiar computations of worldsheet divergences in fixed classical backgrounds 
(for reviews, see \cite{ct,Tseytlin_rev}) one is accustomed to performing $\alpha'$-expansions. It is thus obvious that, 
if following a more conventional route, one would need to follow up with a resummation of leading logarithmic divergences in order to arrive at (\ref{diskdiv}). (Such a resummation has not been performed in \cite{hk}, which is one of the reasons why the implementation of the Fischler-Susskind mechanism in the D0-brane recoil setting has not been elucidated in that publication.)


\section{Strings in a quantum cosmological space}

Having displayed the similarities between the problems induced in string perturbation theory by D0-brane recoil and by gravitational backreaction in a totally compact cosmological space, and having presented the worldline treatment of the D0-brane recoil problem, it is natural to look for a similar solution to the cosmological case. As we shall see in this and the next section, explicitly quantizing a set of large wavelength modes of the background spacetime indeed allows to construct amplitudes with appropriate finiteness properties. Nevertheless, one encounters a considerably greater amount of technical difficulties, compared to the recoil case. 
The reason essentially lies in the well-known conceptual problems arising from ambiguities of gauge-fixing in canonical quantization of gravity, as well as defining observables in general curved space-times. 
With regard to these problems, we shall take a pragmatic approach that, without addressing them in full generality, we can take a particular prescription for quantization that seems appealing to us, and then examine the structure of the amplitudes. In this section, we shall present a very rough sketch of the (potentially divergent) cut-off-dependent terms in the amplitudes, and compare the cut-off-dependent terms arising on spherical worldsheets due to our modification of the formalism with those descending from the torus divergence (\ref{cosmdiv}). In the next section, we shall give a much more detailed technical account discussing a number of subtleties arising in our formalism.

We have seen that introducing an explicit path integral over D0-brane worldlines automatically cancels the divergences associated with D0-brane recoil. Similarly, we shall attempt to integrate over a set of long wavelength modes of the cosmological background to deal with the gravitational backreaction. Which particular modes should be included? The recoil divergence in (\ref{annulus}) came from inserting vertex operators corresponding to the displacement modes of the D0-brane. We hence included a path integral over displacement modes in our reformulation of string perturbation theory. Similarly, the gravitational back-reaction divergences in (\ref{cosmdiv}) come from the zeroth (uniform) Kaluza-Klein modes of the background fields on the spatial torus. It is then natural to consider a modification of string perturbation theory in which a path integral over all such modes is performed. (A different choice would not result in matching cut-off-dependent structures of the kind we observe below.)

One can display the relevant modes explicitly, for the metric, the 2-form and the dilaton:
\beq
\ber{l}
\dsty ds^2=-G_{00}(t)dt^2+2G_{0i}(t)dtdx^i+G_{ij}(t)dx^idx^j,\qquad x^i\in [0,2\pi), \vspace{2mm}\\
\dsty B_{\mu\nu}=B_{\mu\nu}(t),\qquad \Phi=\Phi(t).
\eer
\label{IRmodes}
\eeq
A time-dependent shift of spatial coordinates allows elimination of $G_{0i}$. Subsequently redefining the time variable $t$, one can set $G_{00}$ to a constant value.  
Note that $T=\int dt \sqrt{G_{00}(t)}$ is invariant under time reparametrizations. We shall resort to this gauge in our subsequent treatment.\footnote{The gauge we are using is identical to the synchronous gauge, which is known to be globally ill-defined for general space-times. However, for the homogeneous models we are considering, the gauge can always be reached by the transformations we have just described.} Similarly, the $B_{0i}$ components of the 2-form can be set to zero by a gauge transformation. We are thus essentially dealing with a compactified homogeneous Bianchi I universe.

Quantum dynamics of cosmological spaces whose gravitational degrees of freedom are truncated to a small set of long-wavelength modes has been discussed extensively in the context of Einstein gravity under the name of minisuperspace models. Within this approach, one obviously avoids the well-known ultraviolet problems of Einstein gravity (albeit in an ad hoc manner of truncating the inhomogeneous modes). Nonetheless, a number of inequivalent quantization procedures can be devised (the situation is complicated by the presence of time-reparametrization invariance), and there is no obvious way to decide which quantization procedure should be used \cite{isham,halliwell}. A number of conceptual problems, such as the celebrated `problem of time' arise in relation to these models as well \cite{isham}. As already remarked, we shall take a pragmatic approach to these difficulties by adopting a particularly straightforward quantization prescription and examining its functioning in the context of string theory. More importantly, we shall restrict our attention to the regime of semi-classical dynamics of the background space, where all the quantization subtleties lose much of their importance.

Our purpose is to describe the background field modes given in (\ref{IRmodes}) by an explicit `minisuperspace' path integral, whereas all the other modes are described by strings moving in the background (\ref{IRmodes}), governed by the usual non-linear sigma-model action. We have arrived at this type of formalism by attempting to deal with the issue of infrared divergences arising in conventional string perturbation theory due to gravitational backreaction in a totally compact cosmological space. One could equally well ask whether a minisuperspace quantum cosmology can be given an ultraviolet completion by appending a full set of string modes propagating in it. This would likewise result in a formalism of the type we are currently considering.

In most of the formulas below, we shall suppress the contributions from the 2-form and the dilaton, to make the expressions more compact. These contributions parallel closely those arising from the metric $G_{ij}$, which we shall display explicitly.

With respect to the minisuperspace quantization, we take a conservative approach, which basically amounts to applying the Faddeev-Popov method to deal with the time reparametrization invariance, in the gauge $G_{00}=\cnst, G_{0i}=0$. $G_{00}$ and $G_{0i}$ are thereby completely eliminated from the path integral, except for the constant mode of $G_{00}$. The remaining integral over this constant mode can furthermore be traded for an integral over the duration of experiment $T$, resulting in the following transition amplitudes (cf.\ \cite{marolf,hllwll}):
\beq
\langle G^{(2)}|G^{(1)}\rangle_{V}=\int\limits_{-\infty}^{\infty}\!\! dT\hspace{-5mm}\int\limits_{\ber{l}\ssty G_{ij}(0)=G^{(1)}_{ij}\vspace{-0.5mm}\\ \ssty G_{ij}(T)=G^{(2)}_{ij}\eer}\hspace{-5mm}\D G_{ij}(t)\, e^{iS_{\rm ms}}\int \D X e^{-S_X} \prod\{g_s V^{(n)}\}.
\label{Marolf_string}
\eeq
The square of the absolute value of this amplitude is expected to describe the probability for the quantum universe to have two cross-sections with spatial metrics $G^{(1)}$ and $G^{(2)}$ (and the corresponding values of the 2-form and the dilaton, suppressed in our formulas, as explained above), in the presence of quantum strings described by vertex operators $V^{(n)}$. Note that the vertex operators are `inherited' from the usual Minkowski space S-matrix formulas and are, strictly speaking, not adequate under our present circumstances. 
We shall nevertheless keep them as they are for a moment, and give a more accurate (even if involved) treatment appealing to 
finite time transition amplitudes in the subsequent section. (There  may be situations, for example when the background universe expands to a macroscopic size, in which the usual vertex operator representation of strings becomes nearly exact.) 
In the above formula $S_{\rm ms}$ and $S_X$ are the minisuperspace and the worldsheet actions, which we shall discuss below.

As we have remarked above, the minisuperspace part of the integral (\ref{Marolf_string}) can be understood by starting from a usual quantum-mechanical transition amplitude between the initial and final states of the background (\ref{IRmodes}) and then applying the Faddeev-Popov method to the gauge symmetries that respect the form of the ansatz (\ref{IRmodes}). The integral over the constant (gauge-invariant) mode of $G_{00}$ is traded for an integral over $T$, after which $G_{00}$ and $G_{0i}$ can be replaced by 1 and 0, respectively, in all the expressions in the integrand. (Possible Faddeev-Popov determinant will not affect our considerations, as we shall only need the leading semi-classical approximation to the minisuperspace path integral.) Note that, as a result of the integration over $T$, the `transition amplitude' does not depend on time, which is indeed how it should be in accordance with time reparametrization invariance. We are not talking about the amplitude for a certain change in the state of the system with respect to an external time. Rather, we are talking about an amplitude for the quantum space-time to have two cross sections with specified properties (see, e.g., \cite{HH}).

We shall now specify the actions appearing in  (\ref{Marolf_string}). $S_X$ is simply the conformal gauge non-linear sigma model action in the background  (\ref{IRmodes}) with $G_{00}=1$ and $G_{0i}=0$. 
Explicitly
\beq
S_X=\frac1{4\pi\alpha'}\int d^2\sigma g^{1/2}\left[\left(g^{ab}G_{\mu\nu}(X)+i\epsilon^{ab}B_{\mu\nu}(X)\right)\del_aX^\mu\del_bX^\nu+\alpha'R^{(2)}\Phi(X)\right].
\label{SX}
\eeq
Determination of the minisuperspace action $S_{\rm ms}$ is in principle involved, and should most likely be done order by order in perturbation theory, precisely in a manner that ensures finiteness of the amplitudes. Similarly to the use of the free particle action in the lowest order treatment of the recoil problem, one can start with the Einstein-Hilbert action reduced to (0+1) dimensions (together with the corresponding actions for the 2-form and the dilaton), and observe that it does result in an appropriate cancellation of the (nearly-divergent) cut-off-dependent terms at the lowest non-trivial order. (Higher-order extensions lie outside the scope of this article.) Again, the general form of the relevant dimensionally reduced action can be read off (8.4.2) of \cite{polch}. What is important for us (and what will be explicitly used in our derivations) is that $S_{\rm ms}$ is simply the low-energy bosonic string action with the ansatz (\ref{IRmodes}) substituted in it. (The minisuperspace action for toroidal cosmological spaces has been discussed in \cite{Misner}.)

It is important to keep in mind that (having descended from the Einstein-Hilbert action) $S_{\rm ms}$ comes with a positive power of the Planck mass. We shall focus on the regime when $G_{ij}$ is below the string scale,  which automatically implies that $S_{\rm ms}$ is in a semi-classical regime (and the fluctuations of the background fields are perturbatively suppressed, due to the smallness of the ratio of the string scale to the Planck scale). We hence apply a semiclassical treatment to the minisuperspace part of the path integral in (\ref{Marolf_string}), expanding $G$ around the classical solution $\uG$ to $S_{\rm ms}$ connecting $G^{(1)}$ and  $G^{(2)}$ (and the corresponding expansions for $B$ and $\Phi$, which we do not write explicitly):
\beq
G_{ij}(t)=\uG_{ij}(t)+\gamma_{ij}(t),\qquad\frac{\de S_{\rm ms}}{\de G_{ij}}\Bigg|_{G=\uG}=0,\qquad \uG(0)=G^{(1)},\qquad  \uG(T)=G^{(2)}.
\label{msclass}
\eeq
As the fluctuations of $G$ are suppressed, we expand all structures in (\ref{Marolf_string}) in powers of $\gamma$, retaining only the Gaussian part in $S_{\rm ms}$:
\beq
\langle G^{(2)}|G^{(1)}\rangle_{V}=\int\limits_{-\infty}^{\infty}dTe^{i{\bar S}_{\rm ms}}\int \D\gamma_{ij}(t) e^{iS_{\rm msGauss}(\gamma)}\int \D X e^{-{\bar S}_{X}-\delta S_X} \prod\{g_s V^{(n)}\},
\label{Marolf_expand}
\eeq
where ${\bar S}_{\rm ms}$ is the minisuperspace action evaluated in the background $\uG$, $S_{\rm msGauss}$ is the quadratic part of the expansion of $S_{\rm ms}$ in $\gamma$, ${\bar S}_{X}$ is the worldsheet action in the classical background $\uG$ (independent of $\gamma$), $\delta S_X$ is the remainder of the worldsheet action, which will be treated perturbatively.

Similarly, the integral over $T$ has to be treated by the saddle point method. The saddle point value, $\bar T$, defined by
\beq
\frac{\del{\bar S}_{\rm ms}}{\del T}\Bigg|_{T=\bar T}=0
\label{saddleT}
\eeq
simply corresponds to the length of the time interval for which the solution (\ref{msclass}) to the spatial components of Einstein's equations (obtained by varying the gauge-fixed minisuperspace action $S_{\rm ms}$) subject to boundary conditions given by $G^{(1)}$ and $G^{(2)}$, also satisfies the temporal components of Einstein's equations. Indeed, by the Hamilton-Jacobi equation, $\del{\bar S}_{\rm ms}/\del T=-H$, with $H$ being the canonical Hamiltonian, and the vanishing of the canonical Hamiltonian for a gravitational system is known to imply the Wheeler-DeWitt constraint, which is in turn equivalent to the temporal component of Einstein's equations (see, e.g., \cite{halliwell}). Physically, the fact that the integral ({\ref{Marolf_expand}) is dominated by a saddle point in $T$ is thus a mere reflection of the fact that in a semi-classical regime, the minisuperspace evolution should be dominated by precisely the classical trajectory connecting $G^{(1)}$ and $G^{(2)}$ (with the proper time needed for this process determined by the equations of motion). Note that, after all the deviations from the classical solution connecting $G^{(1)}$ and $G^{(2)}$ are treated perturbatively, the path integrals separate into the Gaussian quasi-classical minisuperspace integration and computations in the (interacting) CFT describing strings\footnote{Conformal invariance, strictly speaking, requires that the background satisfies $\alpha'$-corrected equations of motion, rather than the Einstein equations we have imposed on $\uG$. The difference is unimportant, however, in the low-curvature regime that we shall focus on.} moving in the classical background $(\uG,\bar T)$.

We shall further concentrate on amplitudes for those choices of  $G^{(1)}$ and $G^{(2)}$ for which the classical trajectory dominating the minisuperspace path integral has curvatures low in units of the string scale and the dominant extent of the time interval $\bar T$ is large compared to the string scale.\footnote{Note that the classical equations of motion are completely scale-free, since the Planck mass only appears as an overall prefactor of the action. Hence, the scales for a classical solution are set by the boundary values $G^{(1)}$ and $G^{(2)}$.} In this regime, the usual sigma-model perturbative techniques for the worldsheet path integral are applicable, which will simplify our subsequent analysis. It would have been very interesting to speculate (also in view of phenomenological applications) whether our formulation could be extended to higher curvatures. This would require a fully non-perturbative treatment of the sigma-model (which is presumably only possible numerically), and a suitable definition of the minisuperspace action based on exact sigma-model $\beta$-functions. It is likely that many of the structures we are presenting in this paper will survive in such a setting, but we will presently adopt the modest approach of treating the sigma-model perturbatively. (That we have focused on a semi-classical regime of the uniform background modes by no means implies that we have eliminated all non-trivial stringy physics. Indeed, the worldsheet path integral in (\ref{Marolf_string}) will describe essentially stringy processes, for example, winding string annihilation, which one would not be able to deal with in the context of effective low-energy field theories.)

We shall now give a very crude sketch (to be refined in the next section) of how new cut-off-dependent terms emerge on a sphere worldsheet due to the minisuperspace path integral we have introduced in (\ref{Marolf_string}), and see that their structure matches the contributions from the modular integration on the torus worldsheet. (It is evident that some ultraviolet divergences must be present for any background  configurations in the path integral that do not satisfy the classical equations of motion, but the path integral combines them in a non-trivial way, as our subsequent derivations will show. The deviations from the standard perturbative string theory are, however, suppressed by the large prefactor of $S_{\rm ms}$, which restricts deviations from the classical configuration of the background in the path integral.)

We first examine how the plumbing fixture formula for the modular integration on the torus worldsheet works in our new setting. The modifications must necessarily be minimal, since the torus amplitude already has two extra powers of $g_s$ from genus 1 worldsheet (relatively to the leading order contribution), hence $\delta S_X$ can be neglected altogether (as the fluctuations of $G$ are perturbatively suppressed in $g_s$). The integrals over $\gamma$ and $T$ then become trivial, yielding overall pre-factors that we shall omit in the formulas below. To extract the contributions from degenerating handles, one simply needs to apply the plumbing fixture construction to the CFT describing strings moving in the 
classical background $(\uG, \bar T)$. The derivation of the plumbing fixture construction relies only on conformal invariance and is applicable to interacting worldsheet theories just as much as for the usual Minkowski case. 
The general discussion can be found in   \cite{polchinski-fischler-susskind} and \cite{polch} (see sections 9.4 and 9.5 of \cite{polch} and formulas (9.4.7), (9.5.2) in particular). The torus amplitude is expressed through sphere amplitudes with additional vertex operator insertions:
\beq
\langle G^{(2)}|G^{(1)}\rangle_{V}^{\mbox{\tiny (torus)}}\sim\int \D X e^{-{\bar S}_{X}} \sum_{ab}\int \frac{dq\,d\bar q}{q\bar q}q^{h_a-1}
\bar q^{\tilde h_a-1}{\cal G}^{ab} V_a V_b\prod\{g_s V^{(n)}\},
\eeq
where the sum runs over a complete set of local operators $V_a$ (in the integrated form) with conformal weights $(h_a,\tilde h_a)$, ${\cal G}^{ab}$ is the inverse metric obtained from 2-point functions on a sphere, and we have not kept track of the ghost insertions (which should work in the standard way according to the topology of each given worldsheet). Cutting off the $q$-integral at $|q|=\eps$ produces the following contribution
\beq
\langle G^{(2)}|G^{(1)}\rangle_{V}^{\mbox{\tiny (torus;div)}}\sim\int \D X e^{-{\bar S}_{X}} \sum_{ab}\frac{1-\eps^{2(h_a-1)}}{h_a-1}{\cal G}^{ab} V_a V_b\prod\{g_s V^{(n)}\}
\label{mstorusdiv}
\eeq
(where we have made use of the equality $h_a=\tilde h_a=h_b=\tilde h_b$, following from conformal invariance, and the upper bound of the $|q|$-integration has been set to 1, as setting it to any other value would only yield finite contributions). Because the spectrum of $h_a-1$ extends all the way down to very small values (becoming more and more dense as $\bar T$ increases), and because of the presence of $h_a-1$ in the denominator, dangerous dependences on the cut-off will arise (they would become actual divergences on an infinite time interval). This could be seen explicitly for the flat space-time computation of (\ref{cosmdiv}).

We can now examine the next-to-leading order corrections to the sphere amplitude and observe that a new, nearly divergent, cut-off-dependent term (having no counterpart in the usual string perturbation theory) arises due to the presence of the minisuperspace path integral, and its structure is similar to (\ref{mstorusdiv}). At next-to-leading order, $\de S_X$ can give contributions to the amplitude. Expanding up to second order in $\de S_X$, we obtain:
\beq
\begin{array}{l}
\dsty\langle G^{(2)}|G^{(1)}\rangle_{V}^{\mbox{\tiny (sphere;NLO)}}\sim\int\limits_{-\infty}^{\infty}dTe^{i{\bar S}_{\rm ms}}\int \D\gamma_{ij}(t) e^{iS_{\rm msGauss}(\gamma)}\int \D X e^{-{\bar S}_{X}}\prod\{g_s V^{(n)}\}\vspace{2mm}\\
\dsty\hspace{1cm}\times\left(\int d^2\s_1d^2\s_1\gamma_{ij}(X^0(
\s_1))\del X^i(\s_1)\bar\del X^j(\s_1)\gamma_{kl}(X^0(
\s_2))\del X^k(\s_2)\bar\del X^l(\s_2)\right).
\end{array}
\label{msvertexexpand}
\eeq
Taking the Gaussian integral\footnote{To take this integral one needs to disentangle the functional integral over $\gamma_{ij}$ from the sigma model functional integration over $X^{0}(\sigma)$. 
To do that we use a trick whose essence is in using the identity $\gamma_{ij}(X^{0}(\sigma))=\int dt \gamma_{ij}(t) \delta(t-X^{0}(\sigma))$.  See \cite{thesis} for details. } over $\gamma$ 
merely contracts the two instances of $\gamma$, replacing them by a Green's function $\Gamma_{ij,kl}$. After extracting the leading order contribution from the saddle point in $T$ (quadratic fluctuations in $T$ may in principle contribute at our order, however, their contribution is tame, and we are omitting it here to discuss it more carefully in the next technical section), one ends up with:
\beq
\begin{array}{l}
\dsty\langle G^{(2)}|G^{(1)}\rangle_{V}^{\mbox{\tiny (sphere;NLO)}}\sim\int \D X e^{-{\bar S}_{X}}\prod\{g_s V^{(n)}\}\vspace{2mm}\\
\dsty\hspace{1cm}\times\left(\int d^2\s_1d^2\s_1\Gamma_{ij,kl}(X^0(
\s_1),X^0(\s_2))\del X^i(\s_1)\bar\del X^j(\s_1)\del X^k(\s_2)\bar\del X^l(\s_2)\right).
\end{array}
\label{Greendiv}
\eeq
It should be possible to expand the Green's function $\Gamma_{ij,kl}$ in terms of eigenfunctions $\Gamma^a_{ij}$ of the kernel of $S_{\rm msGauss}$ and the corresponding eigenvalues $\eta_a$:
\beq
 \Gamma_{ij,kl}(t_1,t_2)=\sum\limits_a \frac{\Gamma^a_{ij}(t_1)\Gamma^a_{kl}(t_2)}{\eta_a}.
\eeq
This results in the emergence of the operators 
\beq
{\cal V}^a\equiv \Gamma ^a_{ij}(X^0)\del X^i\bar\del X^j.
\eeq
Equation (\ref{Greendiv}) now takes the form
\beq
\langle G^{(2)}|G^{(1)}\rangle_{V}^{\mbox{\tiny (sphere;NLO)}}\sim\int \D X e^{-{\bar S}_{X}}\prod\{g_s V^{(n)}\}\sum\limits_a \int d\s_1 d\s_2 \frac{{\cal V}^a(\s_1){\cal V}^a(\s_2)}{\eta_a}.
\label{Vdiv}
\eeq
It is important to realize that ${\cal V}^a$ are composite local operators appearing as na\"\i ve insertions in the path integrals and are singular without proper renormalization. We shall study the renormalization properties of these operators in the next section with due care, only to discover that they are multiplicatively renormalized with anomalous dimensions proportional to $\eta_a$. Hence, their renormalization can be expressed (up to coefficients) as
\beq
{\cal V}^a= \de^{\eta_a} V^a,
\label{Vrenorm}
\eeq
where $V^a$ are finite and $\de$ is an ultraviolet cut-off on the worldsheet. Furthermore, the above expression indicates that $V^a$ have definite conformal weights $h_a$ and $\tilde h_a$ equal to 1 (from the explicit worldsheet derivative) plus a term proportional to $\eta_a$. (All of this should hardly be surprising, since ${\cal V}^a$ contain eigenfunctions of the linearized minisuperspace theory, which are related to the eigenfunctions of the graviton Laplacian, and the role of Laplacian eigenfunctions in renormalization of string theory sigma-models at low target space curvature is well-known \cite{friedan}.) Putting everything together, we can rewrite (\ref{Vdiv}) in a form closely reminiscent of (\ref{mstorusdiv}):
\beq
\langle G^{(2)}|G^{(1)}\rangle_{V}^{\mbox{\tiny (sphere;NLO)}}\sim \int \D X e^{-{\bar S}_{X}} \sum\limits_a  \frac{\de^{2(h_a-1)}}{2(h_a-1)}\int\!\! d^2\s_1\,d^2\s_2 \,V ^a(\s_1)V^a(\s_2)\prod\{g_s V^{(n)}\}.
\label{Vdivrenorm}
\eeq

We have thus given a general sketch of how the same structure (insertion of two extra vertex operators together with a propagator-like factor) emerges with the same kind of (nearly divergent) cut-off dependences emerges both in the modular divergence on the torus worldsheet (\ref{mstorusdiv}) and the worldsheet renormalization on the sphere worldsheet (\ref{Vdivrenorm}) after an integral over the long-wavelength modes of the background has been explicitly introduced in (\ref{Marolf_string}). This makes the cancellation of cut-off dependent terms highly plausible (with an appropriate identification of the worldsheet and modular integration cut-offs). Note that even though we have not discussed the overall normalizations (which is a rather subtle question in string perturbation theory), the relative normalizations of the infinite number of terms in the sums of  (\ref{mstorusdiv}) and (\ref{Vdivrenorm}) are identical (the metric ${\cal G}^{ab}$ is trivial, as we shall see in the next section).

Our main purpose for the remainder of the present treatment will be to refine and consolidate the derivation leading from (\ref{Marolf_string}) to (\ref{mstorusdiv}) and (\ref{Vdivrenorm}), which has been presented above in a semi-heuristic manner with a number of important steps omitted. More specifically, the following issues have to be addressed:\vspace{1mm}

\noindent 1) The basic observables of (\ref{Marolf_string}) are defined in a way that intrinsically involves finite time transition amplitudes, and there is no obvious way to avoid this feature (in contrast to the D0-brane recoil problem, where one could define an S-matrix). Dealing with finite time transition amplitudes is a problem known to be difficult in string theory, but we shall be able to isolate these known technical complications from the issues of finiteness. (As we work with large values of $\bar T$, one might expect that the problems associated with defining finite-time string amplitudes should be of secondary importance.)\vspace{1mm}

\noindent $1'$) For dealing with finite time transition amplitudes, the use of vertex operators in (\ref{Marolf_string}) is not strictly speaking appropriate. We shall reformulate the theory in terms of transition amplitudes between fixed boundary loops. Again, this is known to be subtle in string theory and has never been fully worked out, but we shall present a pragmatic treatment sufficient for the finiteness analysis. (Vertex operators may be approximately restored for special classes of amplitudes involving the background universe expanding to macroscopic sizes.)\vspace{1mm}

\noindent 2) We have made strong claims about renormalization properties of local operators involving Laplacian eigenfunctions around (\ref{Vrenorm}). It is important to justify them, in particular, in the context of working on finite time intervals. Similarly, the metric ${\cal G}^{ab}$ in (\ref{mstorusdiv}) needs to be determined.\vspace{1mm}

\noindent 3) Additional considerations are needed to establish the validity of the plumbing fixture construction on a finite time interval.\vspace{1mm}

\noindent 4) We have omitted the contributions in  (\ref{Greendiv}) due to quadratic fluctuations of $T$. It should be explained why they are not important. \vspace{1mm}

\noindent 5) Only spatial components of the worldsheet coordinates appear in (\ref{mstorusdiv}), whereas a priori the plumbing fixture formula contains contributions from all components. We should explain the fate of the temporal components in that context.\vspace{1mm}

\noindent 6) The spectrum of eigenvalues $\eta_a$ on a finite time interval $\bar T$ with Dirichlet boundary conditions is discrete and does not include 0. This implies that the amplitudes we are considering are actually finite even before the cancellation of the cut-off-dependent terms, even though the sums over the Dirichlet-Laplace spectrum become more and more ill-behaved when the value of  $\bar T$ increases. We have to clarify the significance of the cut-off-dependent terms and their cancellation in this context (and to draw parallels to the case of D0-brane recoil, where the scattering times are infinite and cut-off dependences turn into actual divergences).\vspace{1mm}

\noindent 7) Overall signs and normalizations have to be discussed.\vspace{1mm}

\noindent In the next section, we shall present our current understanding of these issues.


\section{Technicalities}

\subsection{Finite time intervals and worldsheets with holes}

Having displayed in the previous section a general sketch of how integrating out fluctuations of the background leads to (nearly divergent) cut-off-dependent terms on worldsheets of a lower genus that have the same structure as the ones coming from modular integration over worldsheets of a higher genus, we shall now supply further technical details closing the gaps in the preceding crude derivation.

Perhaps the most obvious flaw in the discussion of the previous section is that we have been representing the string states that inhabit the compact universe by vertex operators, a language inherited from the usual perturbative string theory S-matrix ansatz. There is no obvious justification for using this language in a situation that lacks asymptotic regions (the space is compact), and hence does not allow for a rigorous notion of `scattering' (in the usual scattering theory sense). Of course, there may be situations when the scattering theory language becomes almost exact (e.g., when the universe expands to sizes much larger than the string scale, which governs the string quanta interactions). Nonetheless, it would be appropriate to give a specification of the quantum amplitudes for our study that does not rely on such approximations, and is applicable under general circumstances.

The compactness of the spatial directions is not the only obstruction to employing scattering amplitudes in our setting. The very definition of the fundamental amplitude (\ref{Marolf_string}), which measures the probability for the quantum universe with a certain set of string excitation to have two cross-sections with given spatial properties, relies essentially on finite time transition amplitudes (with an integral over the duration of the experiment $T$ mandated by the time-reparametrization invariance). It is rather unnatural to try to hybridize\footnote{Note that, for the D0-brane recoil case, one could treat the D0-brane in the language of scattering theory, and define scattering amplitudes for the combined system of D0-brane and strings. We are not aware of similar observables in the case of quantum cosmology, and the minisuperspace part of our fundamental amplitudes (\ref{Marolf_string}) is certainly inherently different from scattering theory amplitudes.} this finite-time transition amplitude with the vertex operator structures inherited from scattering theory (which implies an infinite duration of the experiment).

One could take a step back and envisage a first principles picture of how strings would propagate in our quantized cosmological space-times. A priori, one has a string worldsheet (possibly, with a number of disconnected components)  continuously evolving in the background space-time (with an explicit path integral summation over different worldsheets and different background geometries). The amplitudes of the type given by (\ref{Marolf_string})  measure probabilities for this system to have two spatial cross-sections with given properties. Naturally, the hypersurface defining these cross-sections would cut the string worldsheets open as well, leaving a certain number of boundary loops on the two spatial cross-section of the quantum space-time, see Fig.~\ref{wscut}. This picture leads one to consider (instead of using,  in the manner of (\ref{Marolf_string}), the vertex operators and the corresponding notions of scattering theory) amplitudes specifying explicit string configurations (worldsheet boundary loops) at the two spatial cross-sections of the quantum universe. String amplitudes between given configurations at finite times are known as off-shell string amplitudes, and it is this kind of amplitudes that should appear in combination with minisuperspace path integrals in a general fundamental framework.
\begin{figure}[t]
\centering
\includegraphics[width=6cm, bb=200 560 360 700]{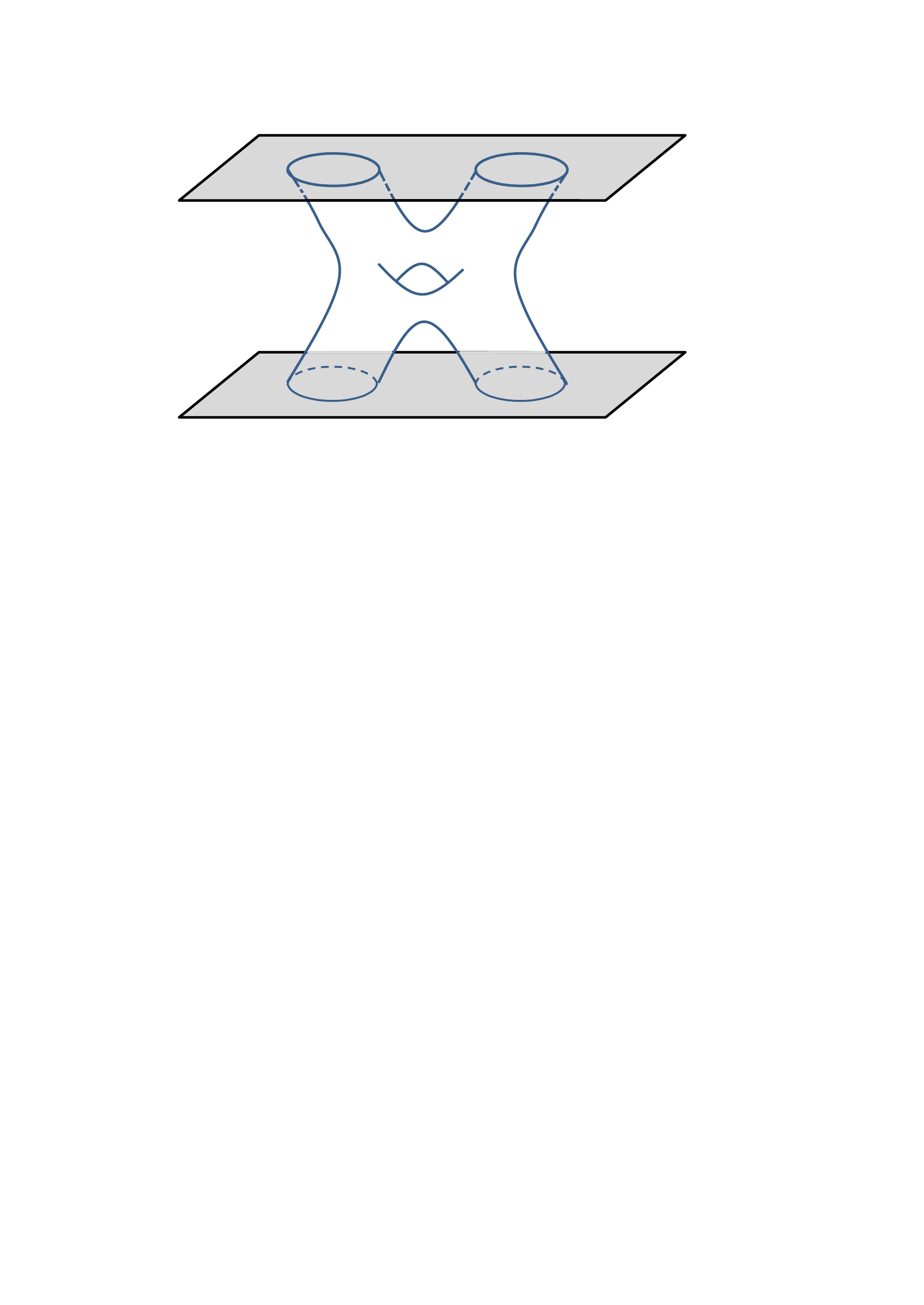}
\caption{A string worldsheet cut open by two observation hypersurfaces.}
\label{wscut}
\end{figure}

Off-shell string amplitudes (for worldsheets with holes attached to given boundary loops) have been discussed extensively (though with fundamental unresolved problems remaining) in the context of Minkowski space-time perturbative string theory, e.g., in \cite{CMNP}. For a general amplitude of this sort, (2.2) in the first paper in \cite{CMNP} gives the following expression 
(after the Gaussian worldsheet path integral specific to the Minkowski target space-time has been performed):
\beq
G(\ell_1,\cdots,\ell_n)=\left(\prod\limits_{i=1}^n\int_{Diff(S^1)}d\Sigma_i\right)\int[d\tau]\left(\mbox{det}'P_1^\dagger P_1\right)^{1/2}(\mbox{det}_{\mbox{\tiny(Dir)}}[-\nabla^2])^{-13}e^{-S_{cl}}.
\label{Mink_loop}
\eeq
Here, one considers an amplitude for the worldsheet to end on $n$ given loops in space-time ($\ell_1,\ldots,\ell_n$). The first integral is over all reparametrizations of the loops, i.e., all the ways to glue the boundary of the worldsheet to the loops (dependence on the parametrization of the loops in the amplitudes would have been unphysical and, indeed, it is eliminated by such integration). Further, $[d\tau]\left(\mbox{det}'P_1^\dagger P_1\right)^{1/2}$ represents the integration over moduli and Fadeev-Popov measure arising from gauge-fixing of the worldsheet metric path integral. Finally, the remaining two factors result from Gaussian integration over the worldsheets, and $S_{cl}$ is the action of the minimal area configuration connecting the loops $\ell_1,\ldots,\ell_n$. Note that, for a given way to glue the boundary worldsheet to the loops $\ell_1$ to $\ell_n$, the worldsheet path integral is simply subject to inhomogeneous Dirichlet boundary conditions (specifying the values of $X$ at the boundary).

Many problems, both technical and conceptual, exist in relation to the expression (\ref{Mink_loop}). Technically, the integrals over reparametrizations are awkward and intractable. Conceptually, the usual decoupling of negative norm states has not been established for amplitudes (\ref{Mink_loop}), in contrast to the usual case of string theory S-matrix, hence their physical content remains somewhat obscure.

In the context of our present research, one can largely circumnavigate these difficulties.
If one considers contributions from worldsheets of different topologies to the same process and discusses the cancellations of cut-off-dependent terms between them, the boundary loops are the same, and the integrals over their reparametrizations are the same. We can hence simply `freeze' those integrals and discuss cancellations for a given parametrization of the boundary. (If cancellations occur before the integral over reparametrizations has been performed, they will occur in the result of the integration as well.) As to the conceptual issues, one can maintain an optimistic attitude that they can be resolved, and, in any case, they do not have much bearing on the cancellation issue. One can furthermore try to define more S-matrix-like quantities in our context (for example, by waiting till the universe expands and becomes large and almost flat) and, for that kind of quantities, conceptual issues have been resolved (in flat space, at least). For large values of $T$, which will be our main focus here, it should also be possible to treat defining finite-time string amplitudes in some kind of perturbative language with $1/T$ corrections (which is an attractive possibility, but not something we shall immediately pursue here). 

We can now combine the picture of a string worldsheet moving in a cosmological space and being sliced by the observation moment hypersurfaces at time 0 and time $T$, the minisuperspace transition amplitudes between universes of different sizes in the spirit of (\ref{Marolf_string}), and a worldsheet amplitude connecting different boundary loops in the spirit of (\ref{Mink_loop}). Namely, we shall consider an amplitude for the quantum universe to have a slice with spatial metric $G^{(1)}$ and an arrangement of strings described by loops $\ell^{(1)}_m$, and another slice with spatial metric $G^{(2)}$ and an arrangement of strings described by loops $\ell^{(2)}_n$ (again, we are ignoring the B-field and the dilaton, but their contributions closely parallel those coming from the metric):
\beq
\langle G^{(2)},\ell^{(2)}|G^{(1)},\ell^{(1)}\rangle=\int\limits_{-\infty}^{\infty}dT\hspace{-6mm}\int\limits_{\ber{l}\ssty G_{ij}(0)=G^{(1)}_{ij}\vspace{-0.5mm}\\ \ssty G_{ij}(T)=G^{(2)}_{ij}\eer} \hspace{-8mm}\D G_{ij}(t)\, e^{iS_{\rm ms}}\left(\prod\limits_{\ell^{(1)}_m\ell^{(2)}_n}\int_{Diff(S^1)}\right)\int\limits_{\ell^{(1)}_m\ell^{(2)}_n} \D X e^{-S_X},
\label{masterclosed}
\eeq
where the modular integration has been suppressed for brevity. The path integral over $G_{ij}(t)$ is subject to the given boundary conditions at $t=0$ and $t=T$, and the integral over worldsheets is subject to Dirichlet boundary conditions that keep it attached to the boundary loops $\ell^{(1)}_m$ and $\ell^{(2)}_n$. Note that the `slicing' picture that led us to (\ref{masterclosed}) makes it clear that the interior of the worldsheet is confined to the interval $X^0\in (0,T)$. It may not extend outside this interval and may not touch the initial and final moments $0$ and $T$ (for, if it did, it would have been cut by our initial and final moment hypersurfaces and would be properly described by introducing additional initial and final state loops). These specific conditions (interval-valued $X^0$) are distinct\footnote{To the best of our knowledge, this space-time slicing picture and the associated constraints on the range of certain worldsheet embedding coordinates has not appeared in the old considerations of off-shell string amplitudes. Our intuition is that such slicing may play a role in establishing proper unitarity properties of the off-shell amplitudes. Note, however, that introducing interval-valued functions complicates path integrals, so that even integrals with a quadratic action cannot be taken exactly, except for special cases (for example, when they can be obtained by orbifolding from the usual integrals over all functional configurations).} from those involved in (\ref{Mink_loop}). The integral over $X$ is written on a Euclidean worldsheet (with the usual analytic continuation necessary to define the integral over $X^0$ that has a wrong-sign kinetic term). The integral over $G_{ij}$ is written in real time. (These are subtle matters, in principle, but hopefully suitable analytic continuations can be defined.)

The idea now is to `freeze' the integral over boundary reparametrizations in (\ref{masterclosed}) and discuss cancellations between sphere (with holes) and torus (with the same set of holes) contributions in the regime when the background curvatures are small. In this situation, the path integrals in (\ref{masterclosed}) are in a quasi-classical regime and can be treated perturbatively. One expands $G_{ij}$ as in (\ref{msclass}).

The subsequent derivation forms a close parallel to the sketch of the previous section. We have to perform the Gaussian integral over the fluctuations of $G_{ij}$ (denoted $\gamma_{ij}$). 
As in (\ref{msvertexexpand}), one gets the following contribution to the sphere amplitude from expanding the exponential of the worldsheet action (\ref{SX}) up to quadratic order in $\gamma_{ij}$ (and performing the saddle point integration over $T$ at leading order):
\beq
\ber{l}
\dsty \langle G^{(2)},\ell^{(2)}|G^{(1)},\ell^{(1)}\rangle^{\mbox{\scriptsize (quad)}}\sim e^{i\bar S_{\rm ms}}\hspace{-8mm}\int\limits_{\ssty \gamma_{ij}(0)=\gamma_{ij}(T)=0\rule{0mm}{3mm}} \hspace{-8mm}\D \gamma_{ij}(t)\, e^{iS_{\rm ms,\gamma}}\int\limits_{\ell^{(1)}_m\ell^{(2)}_n} \D X e^{-\bar S_X}\vspace{2mm}\\
\dsty\hspace{1cm}\times\left(\int d^2\s_1d^2\s_1\gamma_{ij}(X^0(
\s_1))\del X^i(\s_1)\bar\del X^j(\s_1)\gamma_{kl}(X^0(
\s_2))\del X^k(\s_2)\bar\del X^l(\s_2)\right).
\eer
\label{loopGauss}
\eeq
Here, in order to keep the formula compact, we have not displayed the integrals over worldsheet moduli and reparametrizations of the boundary loops, and overall prefactors. $S_{\rm ms,\gamma}$ is the quadratic part of the minisuperspace action expanded in $\gamma$ (the fluctuations of $G$). $\bar S_X$ is the worldsheet action (\ref{SX}) evaluated in the classical background $(\bar G, \bar T)$ (note, in particular, that it does not depend on $\gamma$ and, at our order of approximation, can be treated as a conformal field theory, since the background $\bar G$ satisfies Einstein's equations). The insertion in the last line is simply $\exp(-(S_X-{\bar S}_X))$, expanded to quadratic order in $\gamma$ (the linear order gives a vanishing contribution).

Performing this integral at the lowest order in the quasiclassical expansion simply results in a Green function contracting  the two $\gamma$'s in the integrand. Before displaying this contraction explicitly, we should have a closer look at the structure of the Gaussian action $S_{\rm ms,\gamma}$. Since $S_{\rm ms}$ is simply a restriction of the Einstein-Hilbert action to the minisuperspace configurations (\ref{IRmodes}), $S_{\rm ms,\gamma}$ can be obtained by restricting the linearized Einstein-Hilbert action to the same class of configurations. The linearized Einstein-Hilbert action can be read off \cite{linear,cd,hamber} as\footnote{For a classic treatment of gravity linearization, see \cite{Lichnerowicz}.}
\beq
S_{\rm EH,\gamma}\sim -\frac{1}{g_s^2}\int dx\kappa^{\mu\nu,\rho\sigma}\gamma_{\mu\nu}\bx_{\rm EH}\gamma_{\rho\sigma},
\label{Slin}
\eeq
where $\gamma_{\mu\nu}=G_{\mu\nu}-\uG_{\mu\nu}$ is the metric perturbation (not necessarily restricted to the mini\-superspace modes), and
\beq
\bx_{\rm EH}\gamma_{\rho\sigma}\equiv\nabla_\alpha\nabla^\alpha\gamma_{\rho\sigma}+\uG^{\alpha\beta}\nabla_\rho\nabla_\sigma\gamma_{\alpha\beta}-\nabla^\alpha\nabla_\rho\gamma_{\alpha\sigma}-\nabla^\alpha\nabla_\sigma\gamma_{\alpha\rho},
\eeq
where $\nabla$ is the covariant derivative with respect to $\uG$. ($\Box_{\rm EH}$ equals the Lichnerowicz Laplacian, up to a gradient term that drops out in the harmonic gauge, cf. \cite{Lichnerowicz,Ricciflow}.) The `metric' $\kappa$ is defined by
\beq
\kappa^{\mu\nu,\rho\sigma}= \frac14(\uG)^{1/2}\left(\uG^{\mu\rho}\uG^{\nu\sigma}-\frac12\uG^{\mu\nu}\uG^{\rho\sigma}\right).
\label{kappa}
\eeq
Note that (in neglect of the dilaton) $\kappa$ matches the upper left (gravity-gravity) corner of the metric (2.19) in \cite{TseytlinRG} appearing in considerations of non-linear sigma model renormalization. This is not accidental, and will shall use it shortly in our derivations. We have omitted numerical prefactors and powers of $\alpha'$ (which can be in any case restored by dimension counting) in the gravitational action (\ref{Slin}) but displayed the $1/g_s^2$ prefactor (corresponding to the dilaton dependence of the form $\exp(-2\Phi)$ in the full low-energy string theory action). This coupling-dependent prefactor ensures that the subleading corrections to the sphere amplitude we are analyzing here are of the same order in $g_s$ as the leading contributions to the torus amplitude.

Keeping in mind that $S_{\rm ms,\gamma}$ in (\ref{loopGauss}) is simply obtained from $S_{\rm EH,\gamma}$ by substituting the minisuperspace field configuration:
\beq
S_{\rm ms,\gamma}=S_{\rm EH,\gamma}\Big|_{\gamma_{00}=\gamma_{0i}=0,\gamma_{ij}=\gamma_{ij}(t)},
\eeq
we can now straightforwardly perform the Gaussian integral in (\ref{loopGauss}) to obtain (again, we are neglecting the contributions from quadratic fluctuations of $T$, which turn out to be unimportant, and we shall return to them in section 5.4)
\beq
\ber{l}
\dsty \langle G^{(2)},\ell^{(2)}|G^{(1)},\ell^{(1)}\rangle^{\mbox{\scriptsize (quad)}}\sim g_s^2 \int\limits_{\ell^{(1)}_m\ell^{(2)}_n} \D X e^{-\bar S_X}\vspace{2mm}\\
\dsty\hspace{1cm}\times\left(\int d^2\s_1d^2\s_1\Gamma_{ij,kl}(X^0(
\s_1)),X^0(
\s_2))\del X^i(\s_1)\bar\del X^j(\s_1)\del X^k(\s_2)\bar\del X^l(\s_2)\right),
\eer
\label{greenamp}
\eeq
where the $X$-independent prefactor (determinant times classical action) has been omitted, and $\Gamma_{ij,kl}$ is a Green function satisfying
\beq
\int d\tilde t \frac{\delta^2 S_{\rm ms}}{\delta \gamma_{ij}(t)\delta \gamma_{kl}(\tilde t)}\Gamma_{kl,mn}(\tilde t, t')=\de^i_m\de^j_n\delta(t-t'),\qquad \Gamma_{ij,kl}(0,t')=\Gamma_{ij,kl}(T,t')=0.
\eeq
Using (\ref{Slin}), the equation for $\Gamma_{ij,kl}$ can be re-written as
\beq
\bx_{{\rm EH},t}\Gamma_{ij,kl}(t, t')=\kappa^{-1}_{ij,kl}(t')\de(t-t').
\eeq
The factor $g_s^2$ in (\ref{greenamp}) originates from the prefactor of the action (\ref{Slin}). We are only displaying the power of $g_s$ relative to the leading contribution to the same amplitude from the sphere worldsheet (whose overall power of $g_s$ depends on the number of boundary loops). Note that the power of $g_s$ in (\ref{greenamp}) is the same as in the leading contribution to the same amplitude from torus worldsheets.

Naturally, $\Gamma_{ij,kl}$ in  (\ref{greenamp}) can be expanded in terms of Dirichlet-Laplace eigenfunctions $\Gamma^a_{ij}$ satisfying
\beq
\bx_{{\rm EH}}\Gamma^a_{ij}(t)=-\eta_a\Gamma^a_{ij}(t),\qquad\Gamma^a_{ij}(0)=\Gamma^a_{ij}(T)=0,\qquad
\int\limits_0^T dt \,\kappa^{ij,kl} \Gamma^a_{ij}(t)\Gamma^b_{kl}(t)=\de_{ab},
\label{eigena}
\eeq
as
\beq
\Gamma_{ij,kl}(t, t')=\sum_a \frac{1}{-\eta_a}\Gamma^a_{ij}(t)\Gamma^a_{kl}(t').
\eeq
We hence arrive, in a direct parallel to (\ref{Vdiv}), at the following expression:
\beq
 \langle G^{(2)},\ell^{(2)}|G^{(1)},\ell^{(1)}\rangle^{\mbox{\scriptsize (quad)}}
\sim g_s^2\int\limits_{\ell^{(1)}_m\ell^{(2)}_n} \D X e^{-\bar S_X}\int d^2\s_1d^2\s_1\sum_a
\frac{{\cal V}^a(\s_1){\cal V}^a(\s_2)}{\eta_a},
\label{eigenamp}
\eeq
where
\beq
{\cal V}^a(\s)=\Gamma^a_{ij}(X(\sigma))\del X^i (\sigma)\bar\del X^j(\sigma).
\label{Vgamma}
\eeq
We have thus seen that integrating out fluctuations of the background in the context of the amplitude (\ref{masterclosed}) leads to the emergence of bilocal insertions made of (unrenormalized) vertex operators ${\cal V}^a$. Exploring renormalization of these operators (within the worldsheet CFT with the action $\bar S_X$ defined in the classical background $(\uG, \bar T)$) will reveal, as we alluded to in the sketch in the previous section, nearly-divergent cut-off-dependent expressions of exactly the same structure as those coming from the modular integration (which is what we are aiming to cancel). Note that $\Gamma^a_{ij}(t)$ simply form a subset of Dirichlet eigenfunctions of the linearized gravity Laplacian $\bx_{{\rm EH}}$ that do not depend on spatial coordinates. Investigating renormalization properties of the operators ${\cal V}^a$ will now reveal the cut-off dependences present in (\ref{eigenamp}). It is known that Laplacian eigenfunctions play a special role in non-linear sigma-model renormalization, and we shall now proceed to show that these special properties result in a simple multiplicative renormalization of the operators defined in (\ref{Vgamma}).

\subsection{Conformal dimensions and Dirichlet-Laplace eigenfunctions}

One remaining step in analyzing the cut-off dependences appearing on the sphere worldsheet due to the introduction of a minisuperspace path integral in (\ref{masterclosed}) is to investigate the renormalization properties\footnote{Some systematic considerations of conformal properties of curved space-time vertex operators, albeit without boundaries and without emphasis on the massless sector, are given in \cite{JMW}.} of the composite operators (\ref{Vgamma}) within a worldsheet CFT defined by the non-linear sigma-model in the background $\uG$. Then, the cut-off dependences have to be plugged back into (\ref{eigenamp}). 

We shall proceed (slightly more generally) examining renormalization of the following composite operator inserted on the worldsheet:
\beq
{\cal V}=\int d\sigma\,h_{\mu\nu}(X(\sigma))\del X^\mu(\sigma)\bar\del X^\nu(\sigma).
\label{Vdef}
\eeq
It is convenient to note that
\beq
\int\limits_{\ell^{(1)}_i\ell^{(2)}_j} \D X e^{-\bar S_X} {\cal V} = -\int dx h_{\mu\nu}(x) \frac{\de}{\de \bar G_{\mu\nu}(x)}\int\limits_{\ell^{(1)}_i\ell^{(2)}_j} \D X e^{- \bar S_X}.
\label{diffamp}
\eeq
The amplitude without insertions ($\int \D X e^{-\bar S_X}$) can be treated by the methods usually applied to computing the effective worldsheet action in curved backgrounds (for reviews, see \cite{ct,Tseytlin_rev}). In this approach, one employs a semiclassical approximation by expanding
\beq
X=\uX+\xi,
\label{quasicl}
\eeq
where $\uX$ extremizes $\bar S_X$ (written in the background $\uG$) subject to the boundary conditions attaching the worldsheet to the loops $\ell^{(1)}_i$ and $\ell^{(2)}_j$.
The general idea behind the usual treatment is that $\xi$ as defined by (\ref{quasicl}) does not have convenient transformation properties under target space-time diffeomorphisms. One then trades $\xi$ for another variable $\eta$, which gives components of the (initial) tangent vector to the geodesic connecting $\uX$ and $X$. This substitution reorganizes the Taylor expansions in powers of $\xi$ in a way featuring only covariant quantities. One then changes to vielbein components of $\eta$, which trivializes the kinetic term. The details are given in \cite{ct}, and the resulting covariantized expansions for the worldsheet action are given by (3.19), (3.23) and (3.26) of \cite{ct}. Note that the terms linear in $\eta$ automatically vanish because our background configuration of the worldsheet satisfies the equations of motion. Note also that we are not dealing with the effective action in our present treatment, but are simply inserting the covariantized expansions of \cite{ct} into (\ref{diffamp}), expanded around the dominant classical configuration of the worldsheet.

There is a slight difference between the path integrals in \cite{ct} and the ones we are presently facing. Namely, $X^0$ in our case is constrained to lie between 0 and $T$. This difference, however, has only minor computational consequences when the interval $T$ is large. Indeed, the fluctuations of $\xi$ (or $\eta$) are typically of order $\sqrt{\alpha'}$. This means that the limited range of $X^0$ will only substantially affect the path integral where $|\uX^0|\sim\sqrt{\alpha'}$ or $|\uX^0-T|\sim\sqrt{\alpha'}$. Since the worldsheet stretches from $0$ to $T$, such regions will form very narrow bands around the boundaries of the worldsheet, if $T$ is much larger than $\sqrt{\alpha'}$. Everywhere outside these bands, the path integral over $\eta$ can essentially be treated as unconstrained and the usual formulas derived in non-linear sigma models in infinite target space-time will apply.

Before we proceed treating the boundary effects more carefully, let us ignore them for a moment and give a simplistic sketch clarifying how the Laplace operator $\bx_{\rm EH}$ of (\ref{Slin}) emerges in our present problem. After that, we shall repeat the derivation paying more attention to the boundary effects (Dirichlet boundary conditions for space-time fields at the ends of the time interval $[0,T]$ will emerge from this treatment).

Without the boundary effect, the computation of the divergence of  ($\int \D X e^{-\bar S_X}$) is standard \cite{ct} and yields
\beq
\ber{l}
\dsty\left(\int \D X e^{-\bar S_X}\right)^{(div)}= Ae^{-S_{cl}}\, \log\de\,\int \beta_{\mu\nu}(\uX(\s))\del\uX^\mu(\s)\bar\del\uX^\nu(\s)d\sigma\vspace{2mm}\\
\dsty\hspace{4cm}= \log\de\,\int dx\, \beta_{\mu\nu}(x)\frac{\de}{\de \bar G_{\mu\nu}(x)}\left(\int \D X e^{-\bar S_X}\right),
\eer
\label{sigmadi}
\eeq
where $Ae^{-S_{cl}}$ represents the usual action-determinant prefactor arising from the Gaussian integration in the background of $\uX$, and $\beta_{\mu\nu}$ is the gravity part of the usual\footnote{In the usual treatment of non-linear sigma-models, this divergence is renormalized, which introduces a renormalization group flow of the background space-time metric, as in (2.2) of \cite{TseytlinRG}. In our present approach, the strategy is completely different: the divergence is not renormalized, but its contribution is cancelled against the corresponding terms coming from modular integrations over worldsheets of higher genus contributing to the same string amplitude. The background space-time metric thus does not run. Note also that, if a na\"\i ve worldsheet cut-off $\de$ is used instead of the usual dimensional regularization, power-law divergences in $\de$ can be present in addition to the logarithmic divergence. Those power-law divergences can be renormalized away without breaking Weyl invariance, and this treatment is implicit in our formulas. The logarithmic divergence is essential, however, and results in cut-off dependences that have to be cancelled against contributions from higher genus worldsheets.} $\beta$-function that can be read off, e.g., (2.6) of \cite{TseytlinRG}. The second equality in (\ref{sigmadi}) should essentially be understood in the sense of perturbation theory: it expresses the divergence in a subleading part of the amplitude through the (finite) leading part of the same amplitude.

We can now recall the relations between the $\beta$-functions and the Weyl anomaly coefficient functions $\bar\beta$, and the relation between the latter and the variation of the background action (i.e., the classical equations of motion), expressed by (2.4), (2.5) and (2.13) of \cite{TseytlinRG}. Namely,
\beq
\beta_{\mu\nu}=\bar\beta_{\mu\nu}-\nabla_\mu M_\nu-\nabla_\nu M_\mu=(\kappa^{-1})_{\mu\nu,\rho\sigma}\frac{\de S_{\rm EH}}{\de  \bar G_{\rho\sigma}}-\nabla_\mu M_\nu-\nabla_\nu M_\mu,
\label{betabar}
\eeq
where $M_\mu$ is a certain vector depending on the field variables and defined by (2.5) of \cite{TseytlinRG}, and $\kappa^{-1}$ is the inverse of the `metric' $\kappa$ defined by (\ref{kappa}), which coincides with the upper left corner of (2.18) in \cite{TseytlinRG}. (The inverse has to be evaluated, strictly speaking, taking into account the dilaton entries, but we shall again keep that implicit for the sake of brevity.)

Substituting (\ref{betabar}) in (\ref{sigmadi}), we observe that the terms depending on $M_\mu$ drop out (their contribution to $\beta$ looks like a variation of $\bar G$ under a diffeomorphism, and hence their contribution to (\ref{sigmadi}) will vanish by diffeomorphism invariance of the non-linear sigma model path integral). Hence,
\beq
\left(\int \D X e^{-\bar S_X}\right)^{(div)}\hspace{-3mm}= \log\de\,\int dx (\kappa^{-1})_{\mu\nu,\rho\sigma}\frac{\de S_{\rm EH}}{\de  \bar G_{\rho\sigma}(x)}\frac{\de}{\de \bar G_{\mu\nu}(x)}\left(\int \D X e^{-\bar S_X}\right).
\label{sigmadibar}
\eeq
When inserting this expression back in (\ref{diffamp}), it is important to keep in mind that $\de S_{\rm EH}/\de G$ vanishes at $\bar G$ (by definition of classical solutions), and hence the functional derivative in (\ref{diffamp}) must act on $\de S_{\rm EH}/\de G$ in (\ref{sigmadibar}), since acting on anything else, it will yield a vanishing contribution. Hence,
\beq
\left(\int \D X e^{-\bar S_X} {\cal V}\right)^{(div)}\hspace{-5mm}= -\log\de\int dxdx' h_{\tau\zeta}(x') (\kappa^{-1})_{\mu\nu,\rho\sigma}\frac{\de^2 S_{\rm EH}}{\de  \bar G_{\rho\sigma}(x)\bar G_{\tau\zeta}(x')}\frac{\de}{\de \bar G_{\mu\nu}(x)}\left(\int \D X e^{-\bar S_X}\right).
\label{Vdivbar}
\eeq
Or, taking into account the explicit expression (\ref{Slin}) for the variation of $S_{\rm EH}$ due to variations of $\bar G$, one simply obtains
\beq
\left(\int \D X e^{-\bar S_X} {\cal V}\right)^{(div)}\hspace{-3mm}=- \log\de\,\int dx\, \Big(\bx_{\rm EH}h_{\mu\nu}(x)\Big)\frac{\de}{\de \bar G_{\mu\nu}(x)}\left(\int \D X e^{-\bar S_X}\right).
\eeq
If we now assume, as in (\ref{eigena}), that $h_{\mu\nu}(x)$ is an eigenfunction of $\bx_{\rm EH}$ with an eigenvalue $-\eta$, 
\beq
\bx_{\rm EH}h_{\mu\nu}=-\eta h_{\mu\nu}
\label{hlambda}
\eeq
we obtain:
\beq
\left(\int \D X e^{-\bar S_X} {\cal V}\right)^{(div)}\hspace{-3mm}=\log\de^\eta\,\int dx\, h_{\mu\nu}(x)\frac{\de}{\de \bar G_{\mu\nu}(x)}\left(\int \D X e^{-\bar S_X}\right)=\log\de^\eta \int \D X e^{-\bar S_X} {\cal V}.
\label{intdiv}
\eeq
This expression is consistent with $\cal V$ being an operator of conformal dimension $\eta$, and indeed renormalization group resummation will reproduce its multiplicative renormalization as
\beq
{\cal V}=\de^{\alpha'\eta/4} V,
\label{Veta}
\eeq
where $V$ is a finite renormalized operator.\footnote{The factors of $\alpha'$ and $1/4$ have been restored by dimensional analysis and comparison to the flat space result. They can of course be recovered through a more careful direct computation.} This is precisely as claimed in (\ref{Vrenorm}).

We now turn to the additional subtleties introduced in the above derivation by the fact that, in our setting, $X^0$ takes values on a finite interval $(0,T)$. Our main assertion will be that, if $h_{\mu\nu}$ vanishes on the boundaries of the interval $(0,T)$, i.e., satisfies Dirichlet boundary conditions (imposed on space-time fields rather than on the worldsheet), the leading contribution to the renormalization of (\ref{Vdef}) when $T$ is large remains exactly the same as (\ref{Veta}), derived in neglect of the boundary effects.

To be more precise, let us re-examine the divergence of (\ref{sigmadi}), now keeping close attention to the finiteness of the range of values $X^0$ takes. Once again, we construct the classical (minimal area) solution $\uX$ connecting the initial and final boundary loops, and expand as in (\ref{quasicl}). We shall assume that $T\gg\sqrt{\alpha'}$, denote the whole worldsheet as ${\cal M}$, introduce $\Delta T$ such that $T\gg \Delta T\gg \sqrt{\alpha'}$ and define the following regions on the worldsheet:
\beq
{\cal M}':\,\,\Delta T<\uX^0(\s)<T-\Delta T,\qquad \Delta {\cal M}={\cal M}-{\cal M}'.
\label{Mprime}
\eeq
In other words, $\Delta {\cal M}$ represents the part of the worldsheet mapped to the points in space-time close to the observation moments 0 and $T$ (where boundary effects may be substantial) and ${\cal M}'$ represents the part of the worldsheet far away from the moments of observation where the path integral is essentially unaffected by the boundary conditions (note that the fluctuations of $X$ away from $\uX$ are constrained to be of order of $\sqrt{\alpha'}$). Now, in principle, the range of values of $\xi^0(\s)$ satisfies complicated constraints (depending on the precise shape of $\uX^0(\s)$ and originating from the condition $0<X^0<T$). However, for $\s\in{\cal M}'$ these conditions can be safely ignored (since the action factor in the path integral will not allow $\xi$ to extend beyond the values of order $\sqrt{\alpha'}$, and hence it cannot come even close to violating the constraints anyway). With this simplification in mind, we separate the path integral over $\xi$ into integral over the regions ${\cal M}'$ and $\Delta{\cal M}$, and an integral over the boundary separating these two regions (parametrized by $\theta'$). Such separation is always possible in a path integral with a local action (and, indeed, it is the first step in implementing the plumbing fixture construction we have employed elsewhere in this paper). We write
\beq
\int\limits_{\ell^{(1)}_i\ell^{(2)}_j} \D X e^{- \bar S_X}=\int \D\xi(\theta') \int_{ \Delta {\cal M}}\D\xi\, e^{- \bar S^{\Delta{\cal M}}_X[\xi]} \int_{{\cal M}'}{ \D\xi}\, e^{- \bar S^{{\cal M}'}_X[\xi]},
\label{intsep}
\eeq
where $\bar S^{{\cal M}'}_X$ is the non-linear sigma-model Lagrangian integrated over the region ${\cal M}'$ on the worldsheet and similarly $\bar S^{\Delta{\cal M}}_X$ is integrated over the region $\Delta{\cal M}$. Note that (by construction of the regions ${\cal M}'$  and $\Delta{\cal M}$) $\bar S^{{\cal M}'}_X$ only depends on $\uG_{\mu\nu}$ at $\Delta T<x^0<T-\Delta T$ and $\bar S^{{\cal M}'}_X$ only depends on $\uG_{\mu\nu}$ at $0<x^0<\Delta T$ and $T-\Delta T<x^0<T$. Correspondingly, we can split the functional derivative on the right-hand side of (\ref{diffamp}) into two pieces:
\beq
\int dx h_{\mu\nu}(x) \frac{\de}{\de \bar G_{\mu\nu}(x)}=\hspace{-4mm}\int\limits_{\ber{c}\sssty 0<x^0<\Delta T\vspace{-2mm}\\\sssty T-\Delta T<x^0<T\eer} \hspace{-8mm}dx h_{\mu\nu}(x) \frac{\de}{\de \bar G_{\mu\nu}(x)}\,\,+\hspace{-2mm}\int\limits_{\sssty \Delta T<x^0<T-\Delta T\rule{0mm}{4mm}}   \hspace{-8mm}dx h_{\mu\nu}(x) \frac{\de}{\de \bar G_{\mu\nu}(x)}.
\label{divsep}
\eeq
We shall now act with (\ref{divsep}) on (\ref{intsep}) as in (\ref{diffamp}), assuming that, in addition to the Laplace eigenfunction equation (\ref{hlambda}), $h_{\mu\nu}$ satisfies the Dirichlet boundary conditions at the ends of the interval $[0,T]$,
\beq
h_{\mu\nu}(0)=h_{\mu\nu}(T)=0.
\label{hDir}
\eeq
A solution to the Laplace eigenvalue equation (\ref{hlambda}) on the interval $[0,T]$ with a finite value of $\eta$ should perform a finite number of oscillations on the entire interval $[0,T]$ and one should expect its derivative to be of order $1/T$. Hence, if $h_{\mu\nu}$ vanishes at the ends of the interval, within $\Delta T\ll T$ of the end of the interval, $h_{\mu\nu}$ should be very small (of order $\Delta T/T$). We can therefore neglect the first term on the right-hand side of (\ref{divsep}). Acting with the remaining (second) term in (\ref{divsep}) on (\ref{intsep}), we notice that it will commute with the first two integrations ($\int \D\xi(\theta') \int_{ \Delta {\cal M}}\D\xi$) and hit the integral $\int_{{\cal M}'}{ \D\xi}$. Now,
\beq
\int\limits_{\sssty \Delta T<x^0<T-\Delta T\rule{0mm}{4mm}}   \hspace{-8mm}dx h_{\mu\nu}(x) \frac{\de}{\de \bar G_{\mu\nu}(x)}\int_{{\cal M}'}{ \D\xi}\, e^{- \bar S^{{\cal M}'}_X[\xi]}
\label{interior}
\eeq
has precisely the same structure that we have already analyzed in the derivation leading from (\ref{diffamp}) to (\ref{Veta}), namely, an integral over a worldsheed with a boundary, but without constraints on the fluctuations of the embedding coordinate (as emphasized above, in the region ${\cal M}'$ one can effectively drop those constraints). Hence, the divergence of (\ref{interior}) will be given by (\ref{intdiv}). At the end of the day, we conclude that, if the Dirichlet boundary conditions (\ref{hDir}) are imposed on $h_{\mu\nu}$ in addition to the Laplace eigenfunction equation (\ref{hlambda}), at large $T$, the boundary effects do not affect the renormalization of the operator $\cal V$ of (\ref{Vdef}), which remains the same as the result of our preceding `na\"\i ve' derivation given by (\ref{Veta}).

Finally, substituting (\ref{Veta}) into (\ref{eigenamp}), we obtain an explicit form of the cut-off dependences in the string amplitude, analogous to the heuristic result of (\ref{Vdivrenorm}):
\beq
 \langle G^{(2)},\ell^{(2)}|G^{(1)},\ell^{(1)}\rangle^{\mbox{\scriptsize (quad)}}
\sim g_s^2\sum_a
\frac{\de^{\alpha'\eta_a/2}}{\eta_a} \int\limits_{\ell^{(1)}_m\ell^{(2)}_n} \D X e^{-\bar S_X}\int d^2\s_1d^2\s_1V^a(\s_1)V^a(\s_2).
\label{eigenrenorm}
\eeq

\subsection{The plumbing fixture construction on a finite time interval}

Having completed the analysis of renormalization properties of composite operators in the non-linear sigma-model describing strings in the classical background $\uG$, and, as a consequence, having derived an explicit expression for the nearly-divergent cut-off-dependent contributions to the amplitude (\ref{masterclosed}), we shall now proceed reviewing a few relatively minor subtleties arising in the plumbing fixture construction applied to (\ref{masterclosed}).

The plumbing fixture construction is formulated in a few steps. First we cut the torus worldsheet to create a cylinder and represent the original worldsheet path integral on the torus as an integral over values of $X$ on the cylinder and on its boundary. The latter can be rewritten as a sum over a complete set of states on the boundary. Afterwards, two disks with one local operator insertion on each can be used to seal the boundaries (and the local operators are chosen to reproduce precisely the said complete set of states on the boundaries of the cylinder). Subsequently, the path integral becomes that over a sphere worldsheet with two additional local insertions. Finally we relate the modular parameter of the torus to the size of the holes introduced by cutting the torus and use scaling properties of the local operators to give the plumbing fixture representation its final form. (Note that the cutting step applies in any local field theory, and it is only afterwards that the conformal nature of the worldsheet theory is essentially used.) At the end of the day we obtain the following representation (again, boundary reparametrization integral, $T$-integral and Gaussian integration pre-factors from the semi-classical treatment of $G(t)$ are omitted, as we did for the sphere amplitude):
\beq
 \langle G^{(2)},\ell^{(2)}|G^{(1)},\ell^{(1)}\rangle\rule{0mm}{5mm}^{(div,torus)}\hspace{-5mm}\sim g_s^2
\sum_{ab}\int \frac{dq\,d\bar q}{q\bar q}q^{h_a-1}
\bar q^{\tilde h_a-1}{\cal G}_{ab}\hspace{-3mm}\int\limits_{\ell^{(1)}_i\ell^{(2)}_j} \hspace{-3mm}\D X e^{-S_X} \int d\sigma_1 d\sigma_2\ V^a(\sigma_1) V^b(\sigma_2).
\label{plumbing}
\eeq
(${\cal G}^{ab}$ is the 2-point `gluing metric' defined by a sphere amplitude with $V^a$ and $ V^b$ inserted on the sphere, and $h_a$ and $\tilde h_a$ are the conformal weights of $V^a$.) As in (\ref{greenamp}) and (\ref{eigenrenorm}), we only display the power of $g_s$ relative to the leading (sphere worldsheet) contribution to the same amplitude. $g_s^2$ corresponds to the torus worldsheet topology.

We would now like to prove that $V$ defined by (\ref{Vdef}), (\ref{hlambda}), (\ref{Veta}) and (\ref{hDir}) are: (1) definite conformal weight operators and (2) being inserted in the interior of a disk, produce a complete set of states on its boundary. The first statement is essentially proved in the previous section. The conformal weights are determined as a sum of the na\"\i ve dimension (the number of worldsheet derivatives) and the $\eta$-dependent anomalous dimensions appearing after renormalization (\ref{Veta}):
\beq
h=\tilde h = 1+\alpha'\eta/4.
\eeq
We still have to prove that, inserted in the interior of a disk, the operators $V$ produce a complete set of states on its boundary (after the path integration in the interior has been performed). Importantly, since curvatures are small in our setting (by our choice of $G^{(1)}$ and $G^{(2)}$), we need to establish this statement to zeroth order in the quasiclassical expansion of the non-linear sigma model.

To be more precise, we are interested in boundary wave functionals defined by
\beq
\Psi[X(\te)]=\int_{X(\te)} dX e^{-S_X} V,
\label{Psidef}
\eeq
where the worldsheet action is written in the given space-time background and the path integral is performed over all configurations equal $X(\te)$ at the boundary of the disk (parametriz\-ed by $\te$). As explained above, we are computing $\Psi[X(\te)]$ at zeroth order in the space-time curvature. Hence, it should suffice, at our order of approximation, to consider the semi-classical result
\beq
\Psi[X(\te)]= Ae^{-S_X[\uX]} V[\uX],
\label{PsiGauss}
\eeq
where $A$ is the Gaussian determinant factor, and $\uX$ is the classical solution equal to $X(\te)$ at the boundary. Now, if $X(\te)$ extends over a region considerably bigger than $\sqrt{\alpha'}$, then $S_X[\uX]$ is huge, $\Psi[X(\te)]$ is negligible, and we do not need to worry about its precise value. If $X(\te)$ is contained in a region whose size is of order $\sqrt{\alpha'}$, then our background fields vary extremely little in this region (their variation is negligible at zeroth order in space-time curvature) and $\uX$ will simply be given by the corresponding flat space-time expression. In flat space-time, it is known what $\Psi$ is:
\beq
\Psi_{flat}[X(\te)]=h_{\mu\nu}\left(\int d\te X(\te)\right)|1,0,0,\cdots\rangle^\mu|\tilde 1,\tilde 0,\tilde 0,\cdots\rangle^\nu,
\eeq
where $|1,0,0,\cdots\rangle^\mu$ and $|\tilde 1,\tilde 0,\tilde 0,\cdots\rangle^\nu$ represent wave functions of the left- and right-moving oscillators of the string (whose arguments are given by Fourier components of $X(\te)$) with the given occupation numbers, and $h_{\mu\nu}$ is evaluated on the coordinate of the zero-mode of the string $\int d\te X(\te)$. There is one more subtlety: the Gaussian expression (\ref{PsiGauss}) is, strictly speaking, incorrect when $X(\te)$ lies close to the boundary of space-time. However, as explained in the previous section, $h_{\mu\nu}$ very small in that region (due to the Dirichlet boundary conditions and the rate of variation of $h_{\mu\nu}$), so again, $\Psi$ is negligible and the error we are making is irrelevant at our order of approximation. Overall, we obtain the following form of $\Psi$:
\beq
\Psi[X(\te)]=
\left\{
\ber{ll}
\dsty h_{\mu\nu}\left(\int d\te X(\te)\right)|1,0,0,\cdots\rangle^\mu|\tilde 1,\tilde
 0,\tilde 0,\cdots\rangle^\nu\quad &\mbox{if }\forall\te:X^0(\te)\in(0,T)\vspace{1mm}\\
\dsty 0&\mbox{otherwise}.
\eer
\right.
\label{Psibdry}
\eeq
Different arrangements of derivatives of $X$ in $V$ will results in different excitation numbers of the oscillators (however those operators are not of interest to us, since they do not give divergent contributions in the plumbing fixture construction). Overall, completeness of Dirichlet eigenfunctions on the interval $(0,T)$ (and completeness of the oscillator occupation number basis) guarantee the completeness of the set of $\Psi$'s generated by our local operators, and justify the use of the operators $V$ in the plumbing fixture formula (\ref{plumbing}). 

The approximate expression for boundary states (\ref{Psibdry}) generated by local operator insertions on a disk also allows to determine the `gluing metric' ${\cal G}_{ab}$ at our level of precision quite easily. Indeed, a sphere with two vertex operator insertions, $V^a$ and $V^b$, can be constructed by gluing along the boundary two disks with one operator on each. Then the sphere path integral will simply equal $\langle \Psi^a|\Psi^b\rangle$, where the $\Psi$'s are defined by (\ref{Psidef}). Given the usual orthonormality relations between the Dirichlet eigenfunctions and the oscillator eigenstates, one concludes that ${\cal G}_{ab}=\de_{ab}$ in the basis of operators defined by Dirichlet-Laplace eigenstates $h_{\mu\nu}$.

Putting everything together and extracting from (\ref{plumbing}) the (nearly divergent) contribution due to a degenerating handle from the $q$-integral (cut off at $\eps$), we obtain:
\beq
 \langle G^{(2)},\ell^{(2)}|G^{(1)},\ell^{(1)}\rangle^{(div,torus)}\sim g_s^2
\sum_{a}\frac{1-\eps^{\alpha'\eta_a/2}}{\eta_a}
\int\limits_{\ell^{(1)}_i\ell^{(2)}_j} \D X e^{-S_X} \int d\sigma_1 d\sigma_2 V_a(\sigma_1)V_a(\sigma_2).
\label{eigentorus}
\eeq
The cut-off dependent part is manifestly of the same form as (\ref{eigenrenorm}), the contribution on the sphere introduced by including the minisuperspace path integral in (\ref{masterclosed}).

\subsection{Fluctuations of the time interval}

There is a number of subtleties we have omitted from our presentation. We have postponed the discussion until now as it does not have any major bearing on the big picture. One must remember, in particular, that all amplitudes should be integrated over $T$. This integration descends from the one in our fundamental definition of the amplitudes (\ref{masterclosed}), and as alluded to around (\ref{saddleT}) it is dominated by a classical saddle point in the regime we are considering.

The way the saddle point evaluation of the $T$-integral operates is different, however, in the context of the sphere and the torus amplitudes. The torus amplitude already has extra powers of the coupling, and the integration over $T$ has to be considered only up to the leading saddle point contribution, which amounts to replacing $T$ by $\bar T$ of (\ref{saddleT}), and appending some overall pre-factor. The same will hold true for the nearly-divergent contribution to the sphere amplitude given by (\ref{eigenrenorm}) which also has additional powers of the coupling (from the integration of the gravitational fluctuations). In the end, the cut-off-dependent parts of (\ref{eigenrenorm}) and (\ref{eigentorus}) will still cancel each other after the saddle point integral has been performed. (Note that, technically, it is quite important to perform this integral for the torus amplitude since, strictly speaking, only for $T$=$\bar T$ all Einstein's equations are satisfied and the corresponding worldsheet theory is conformal, at our order of approximation, which enables one to apply CFT techniques, such as the plumbing fixture formula.)

However, besides (\ref{loopGauss}) one can consider a term which is at leading (rather than subleading) order with respect to $\gamma(t)$:
\beq
\langle G^{(2)},\ell^{(2)}|G^{(1)},\ell^{(1)}\rangle^{(\mbox{\scriptsize lead},\gamma)}\sim \int dT \, e^{i\bar S_{\rm ms}}\int\limits_{\ell^{(1)}_m\ell^{(2)}_n} \D X e^{-\bar S_X}
\label{looplead}
\eeq
 and then take into account the fluctuations of $T$ around the saddle point value $\bar T$ in that term. This would produce additional corrections to the sphere amplitude, which is what we would presently like to analyze.

To appreciate the absence of trouble in the contributions arising from fluctuations of $T$, it is important to keep in mind that (unlike the fluctuations of $\gamma(t)$) there is only one integration variable involved, hence, the saddle point evaluation will only produce one denominator (which is the inverse of the second derivative of $S_{\rm ms}$ with respect to $T$) rather than a whole series of different denominators, some of which may become small leading to potential divergences as in (\ref{eigenrenorm}). Essentially, one obtains:
\beq
\ber{l}
\dsty\langle G^{(2)},\ell^{(2)}|G^{(1)},\ell^{(1)}\rangle^{(\mbox{\scriptsize lead},\gamma)}\sim e^{i\bar S_{\rm ms}(\bar T)} \int dT \exp\left[i\frac{\del^2 S_{\rm ms}}{\del T^2}\Biggl|_{T=\bar T}(T-\bar T)^2 \right] \vspace{2mm}\\
\dsty\hspace{2cm}\times \int\limits_{\ell^{(1)}_m\ell^{(2)}_n} \D X \exp\left[-\bar S_X(\bar T)+\frac{\del \bar S_{X}}{\del T}\Biggl|_{T=\bar T}(T-\bar T)+\frac{\del^2 \bar S_{X}}{\del T^2}\Biggl|_{T=\bar T}(T-\bar T)^2\right].
\eer
\label{Texpand}
\eeq

Since considering the theory on the time interval $(0,T)$ with $G_{00}=1$ is the same as considering the theory on the time interval $(0,\bar T)$ with $G_{00}=T^2/\bar T^2$, one can write
\beq
\frac{\del}{\del T}= \frac{2T}{\bar T^2}\int\limits_0^{\bar T}dt\,\frac{\de}{\de G_{00}(t)}.
\eeq
Hence, every differentiation (of a reparametrization-invariant functional) with respect to $T$ in (\ref{Texpand}) will effectively produce a factor of $1/\bar T$. Then, ${\del^2 S_{\rm ms}}/{\del T^2}$ will be accompanied with a factor of $1/g_s^2 \bar T^2$ (where $g_s^2$ comes from the prefactor of the action itself). As one expands the second line of (\ref{Texpand}) in a power series in $(T-\bar T)$, all terms proportional to $(T-\bar T)^2$ will be accompanied by $1/\bar T^2$, from the derivatives of $\bar S_X$. At the end of the day, after the saddle point integration over $T$ has been performed, one is left with innocuous contributions of order $g_s^2$, as suggested above. (We are specifying all powers of $g_s$ relatively to the leading contribution to a given amplitude.)

\subsection{Temporal versus spatial contributions in the plumbing fixture formula}

We have seen that the integration over quadratic fluctuations of the spatial metric $\gamma_{ij}(t)$ produces cut-off-dependent terms very similar to those emerging from the plumbing fixture construction, except that the indices of $\del X$ only run over spatial values (and the integration over quadratic fluctiations of $T$ does not produce any essential extra contributions). We will now consider the role of the $\del X^0$-dependent terms in the plumbing fixture formula (\ref{plumbing}) and sketch an argument as to why they should be free from divergences.

A key observation that points to decoupling of the temporal and spatio-temporal components in the plumbing fixture formula (\ref{plumbing}) is that any spatially uniform metric can be brought to the synchronous gauge, as explained under (\ref{IRmodes}). In view of this, one can split each vertex operator of the form (\ref{Vdef}) as
\beq
V_a=\tilde V_a + \Delta V_a,
\eeq
where $\tilde V_a$ is of the same form as $V_a$, except that $h_{00}$ and $h_{0i}$ are set to 0, and $\Delta V_a$ is of the same form as $V_a$, except that $h_{ij}$ are set to 0. Then, because $h_{00}$ and $h_{0i}$ can be set to 0 by a gauge transformation (we shall not pay attention to the possible boundary terms here), it should be possible to write
\beq
\Delta V_a = \int d\s (\bar\nabla_\mu\xi_\nu+\bar\nabla_\nu\xi_\mu)\del X^\mu \bar\del X^\nu,
\eeq
where $\bar\nabla$ is the covariant derivative with respect to the background metric $\bar G$ and $\xi$ is the (linearized) gauge transformation that accomplishes elimination of $h_{00}$  and $h_{0i}$. Because the operator $\Delta V_a$ is constructed from a pure gauge configuration of $h_{\mu\nu}$ it should be BRST-exact and its contribution should drop out from the amplitudes after the integration over worldsheets. One is thus left only with contributions of $\tilde V_a$, which only includes spatial components of the worldsheet embedding, just like the contributions on the sphere due to background fluctuations given by (\ref{eigenamp}).

\subsection{Finiteness, infrared safety and cut-off-dependent terms}

As we have remarked before, the amplitudes we are computing are dominated by a finite value\footnote{That $\bar T$ is finite for generic values of $G^{(1)}$ and $G^{(2)}$ follows from the power law character of the Kasner solutions describing our background. Indeed, $\bar T$ could only diverge at a certain value $G^{(2)}=G^{(2)}_0$ if there existed a classical trajectory approaching $G^{(2)}_0$ at infinity. But  according to the Kasner-dilaton solutions, the eigenvalues of $G$ (treated as a matrix) behave as power laws, hence $G$ cannot have a finite limit at large times. Some formulas pertaining to Kasner-dilaton solutions are given in \cite{kasner1,kasner3, kasner2}.} of the evolution time $T=\bar T$ and the Dirichlet-Laplace spectra appearing in (\ref{eigenrenorm}) and (\ref{eigentorus})  are discrete and do not contain 0. Hence,  (\ref{eigenrenorm}) and (\ref{eigentorus}) are finite separately, and so is, of course, their sum. There are no divergences in the amplitude (and this is ultimately a reflection of the fact that our observables are defined by path integrals on finite time intervals with Dirichlet boundary conditions on background fields).

In principle, one could rest content with the statement of finiteness. This perspective would have been rather short-sighterd, however. Given that we know that the amplitudes would become divergent if $\bar T$ had been artificially sent to infinity, it is appropriate to study their behavior at moderate values of $\bar T$ and see whether any precursors to divergences arise. Considering the limit of large time intervals would also be useful for establishing contact with our earlier work on D0-brane recoil \cite{d0} formulated in terms of (infinite time) scattering observables.

$\bar T$ can be increased in a controlled fashion by moving $G^{(2)}$ along a classical trajectory (starting at $G^{(1)}$) to greater and greater times. As this happens, various divergences will develop in (\ref{eigenrenorm}) and (\ref{eigentorus}), and we shall presently discuss their meaning.

The divergence in (\ref{eigentorus}) comes in two parts. The first and simpler part is
\beq
\sum_{a}\frac{1}{\eta_a}
V_a(\sigma_1)V_a(\sigma_2)=\left\{\Gamma_{kl,mn}(X^0(\s_1),X^0(\s_2))\del X^k\bar\del X^l(\s_1)\del X^m\bar\del X^n(\s_2)\right\}_{ren}
\label{rengreen}
\eeq
(where $\Gamma_{kl,mn}(X^0(\s_1),X^0(\s_2))$ is the Dirichlet-Laplace Green's function, and the subscript $ren$ signifies that the expression is renormalized with respect to the worldsheet path integration, since the operators $V_a$ appearing in the plumbing fixture construction are renormalized). This expression, of course, becomes divergent if $\bar T$ is sent to infinity, and this behavior merely reflects the non-existence of the Dirichlet Green's function on an infinite time interval. The reason why the Dirichlet Green's function appears in our amplitudes (rather than some other Green's function) is that we essentially started with finite-time transition amplitudes, which typically do not have an infinite-time limit. One could try to construct other amplitudes that do have an infinite-time limit (like the S-matrix for scattering dynamics), and that would have changed the kind of Green's functions appearing in perturbative expansion as well. We shall briefly comment on such possibilities below. Be it as it may, the above ``renormalized Green's function'' term in the amplitude is completely innocuous for moderate values of $\bar T$ (say, $10\sqrt{\alpha'}$).

There remains the cut-off-dependent part of (\ref{eigentorus}) which has been the main focus of our considerations:
\beq
\sum_{a}\frac{\eps^{\alpha'\eta_a/2}}{\eta_a}
V_a(\sigma_1)V_a(\sigma_2).
\eeq
This part should cancel the sphere contribution (\ref{eigenrenorm}), with an appropriate identification of the cut-offs $\eps$ and $\de$. Had the cancellation not occured, one would have had additional divergences. Unlike the ``renormalized Green's function'' term considered above, these cut-off-dependent terms exhibit troublesome behavior even for moderate values of $\bar T$ (irrespectively of taking the infinite time limit). The reason is not the magnitude of these terms, but rather their extremely fast dependences on the cut-off. Indeed, the above expression, strictly speaking, goes to 0 when $\eps$ goes to zero, but this smallness will only be achieved for $\eps\ll \exp[-1/\alpha'\eta^*]$, where $\eta^*$ is the smallest Dirichlet-Laplace eigenvalue. Since the smallest eigenvalue goes to 0 with $\bar T$ going to infinity (for example, it would go like $1/\bar T^2$ for free particle motion), we are talking about an exponentially small number, even for moderate values of $\bar T$. If the value of the cut-off-dependent terms crucially depends on whether the worldsheet cut-off is $10^{-100}$ or $2\cdot 10^{-100}$, one is clearly in trouble at the level of any practical computation. The Fishler-Susskind-like cancellation of cut-off-dependent terms between the sphere and the torus contributions to the same amplitude alleviates the issue. There is a clear parallel between the cancellation of these cut-off-dependent terms in our present case, and the cancellation of divergences for (infinite-time) observables of the D0-brane recoil case in \cite{d0}.

To make better contact with the scattering theory considerations, it could be advantageous to define actual `scattering' observables for our present minisuperspace system. Here, `scattering' does not refer to scattering of strings inside the cosmological space, but rather to some sort of amplitudes describing infinite-time evolution of the minisuperspace itself (instead of the amplitudes for the universe to have two given cross-sections, which have been our main focus here). Dynamics of Kasner space-time simplifies at late times, becoming more and more semi-classical, which could give one hope that the late-time simple evolution can be `amputated' yeilding well-defined infinite time amplitudes. It is much more difficult to specify, however, in which sense the dynamics of the string gas filling the universe becomes simple at late times (though it might, if the string gas becomes dilute). We shall not pursue the construction of infinite-time observables here. 

\subsection{Signs and normalizations}

Summing up what has been done so far, we have analyzed transition amplitudes in our formalism and discovered that they are finite (which is basically predetermined by their construction, based on finite time transition amplitudes). Additionally, we have discovered that, even though the limit of zero cut-off exists (for both worldsheet cut-off and modular integration cut-off), the dependences on the cut-offs are dangerously near-singular (and this behavior is a reflection of the infrared divergences on an infinite time interval, associated with gravitational backreaction in a compact cosmological space). We have further seen that the functional forms of these near-singular terms are identical for sphere and torus worldsheets, and hence they can be cancelled in the manner of the Fischler-Susskind mechanism, provided that their overall coefficients match as well. The question of determining these overall coefficients is rather subtle, however, and we shall limit ourselves to a general discussion and making analogies to the D0-brane recoil case.

It is a general feature of string perturbation theory that each amplitude receives contributions from separate path integrals on worldsheets of different genera, and care needs to be taken in choosing path integral normalization coefficients for each genus of the worldsheet. In the usual calculation of string theory S-matrix in Minkowski space-time, relative normalizations are fixed by demanding unitarity. One may also do this indirectly, i.e., by matching the low-energy limit of the string amplitudes to amplitudes derived from the corresponding effective low-energy field theory (which are unitary by construction).

In the previous treatment of D0-brane recoil \cite{d0}, a variant of such indirect approach was applied. The normalizations of the different divergent contributions to the amplitudes were fixed through an appeal to the low-energy DBI action (with some natural assumptions on how such divergent amplitudes should be represented in the DBI language). The result (together with a geometrically inspired relation between the worldsheet and the modular integration cut-offs) displayed an exact cancellation of the two divergent contributions to the scattering amplitude.

One may try to pursue a similar strategy in our present setting, but the logic of such a derivation appears somewhat flawed. Indeed, it is natural to normalize the amplitudes in such a way that the near-singular cut-off dependences exactly cancel and discuss whether the remaining well-behaved part satisfies appropriate unitarity conditions. 

For the D0-brane recoil case, such an approach could be easy to implement, as long as the finite parts of the amplitudes are carefully calculated. Indeed, the unitarity conditions for scattering matrices are well understood and it is easy to check whether they are satisfied or not, at a given order in the string coupling. For our present case, we are not dealing with scattering amplitudes, but rather with the time-reparametrization-invariant amplitudes (\ref{masterclosed}). It should be possible, in principle, to develop an analogue of unitarity conditions for such amplitudes, and then investigate whether the finite part remaining after the cancellation of divergences (\ref{eigenrenorm}) and  (\ref{eigentorus}) satisfies these conditions.

We shall not presently pursue this program in any detail. The fact that the functional form of the cut-off dependences in (\ref{eigenrenorm}) and  (\ref{eigentorus})  is the same (including an infinite number of relative coefficients of the different terms in the sums), and hence their cancellation can be arranged by tuning the overall path integral normalization coefficients, is strongly indicative that our formalism is on the right track. The fact that our defining path integral (\ref{masterclosed}) is explicitly constructed by slicing a continuous space-time process with observation moment hypersurfaces is also intuitively suggestive of unitarity being respected (for unitarity arises in the path integral formalism from gluing path integrals on two adjacent time intervals being the same as the path integral on the combined time interval).


\section{Conclusions}

We have addressed the problem of infrared divergences (indicative of gravitational backreaction) in the worldsheet theory of quantum relativistic strings propagating in a fully compactified space. In order to cure the theory of infinities, we have proposed a modification of the usual string perturbation theory in which a path integral over spatially uniform modes of the background fields is performed (in addition to the usual worldsheet path integral). This formalism is modelled on the related solution of the D0-brane recoil problem (which is in turn a string theory implementation of the usual collective coordinate treatment of soliton recoil). 

The natural quantities to consider in our formalism are transition amplitudes for the universe to have two cross-section with specified spatial properties and string contents. For such amplitudes, the finite proper time required to accomplish the transition serves as an infrared cut-off and resolves the infrared divergences present in the usual theory on an infinite time interval. Nonetheless, one discovers highly pathological cut-off dependences in the amplitudes (inherited from the divergences on an infinite time interval). These cut-off dependences can in turn be cancelled between the lowest genus and the next-to-lowest genus topologies of the worldsheet in the manner of the Fischler-Susskind mechanism.

It would be interesting to re-examine the intriguing phenomenological conjectures made for the dynamics of compact cosmological spaces filled with a string gas in our present setting. (The question of the number of spatial dimensions that can expand to macroscopic sizes is particularly important.) The role of T-duality in our formalism should be equally interesting to investigate (for it should become a symmetry of the amplitude (\ref{masterclosed}), rather than a duality connecting theories living in different background space-times). Even though establishing the validity of our formalism has required performing computations beyond the leading order in the string coupling (and analyzing divergence cancellation in the first subleading order), the physically interesting applications of our formalism we have just mentioned are likely to depend (predominantly) on the results at leading order in the string coupling.\footnote{The stretched worldsheet action for each initial and final string configuration will modify the amplitudes for the background universe to evolve between two states given by (\ref{masterclosed}).} Computationally, this would certainly be a welcome simplification.

\section{Acknowledgments}

We would like to thank Tom Banks, Micha Berkooz, Ben Freivogel, Simeon Hellerman, Rajesh Gopakumar, Donald Marolf, Rob Myers, Joe Polchinski, Ashoke Sen, Joan Simon and Arkady Tseytlin for useful discussions at various stages of this project.  The work of B.C.\ has been supported by the Belgian Federal Science Policy Office through the Interuniversity Attraction Poles IAP VI/11 and P7/37, by FWO-Vlaanderen through project G011410N, and by the Vrije Universiteit Brussel through the Strategic Research Program ``High-Energy Physics.'' The research of O.E.\ has been supported by Ratchadaphisek Sompote Endowment Fund. The research of A.K.\ was supported in part by the STFC grants 
ST/G000514/1 ``String Theory Scotland'' and ST/J000310/1 ``High energy physics at the Tait Institute.''  O.E. would also like to thank the string theory group of Harish-Chandra Research Institute (Allahabad, India) and all participants of Kumbh Mela 2013 for a stimulating environment during early stages of the manuscript preparation.


\begin{thebibliography}{99}
\bibitem{rajaraman} R. Rajaraman, {\it Solitons and instantons}, North Holland, 1982.
\bibitem{christ-lee}   N. H. Christ and T. D.  Lee, {\it Quantum expansion of soliton solutions}, \\
Phys. Rev. {\bf D12} (1975) 1606. 
\bibitem{fs}W.~Fischler and L.~Susskind,
  {\it Dilaton tadpoles, string condensates and scale invariance},
  Phys.\ Lett.\ B {\bf 171} (1986) 383; {\it Dilaton tadpoles, string condensates and scale invariance 2},
  Phys.\ Lett.\ B {\bf 173} (1986) 262.
\bibitem{d0}B.~Craps, O.~Evnin and S.~Nakamura,
  {\it D0-brane recoil revisited},
  JHEP {\bf 0612} (2006) 081
  [\arXiv{hep-th/0609216}].
\bibitem{thesis}O.~Evnin, {\it On quantum interacting embedded geometrical objects of various dimensions}, Caltech Ph.D.\ thesis (2006), available electronically at http://thesis.library.caltech.edu/2495/
\bibitem{localrecoil}B.~Craps, O.~Evnin and S.~Nakamura,
  {\it Local recoil of extended solitons: a string theory example},
  JHEP {\bf 0701} (2007) 050
  [\arXiv{hep-th/0608123}].
\bibitem{bv}R.~H.~Brandenberger and C.~Vafa,
  {\it Superstrings in the early universe},\\
  Nucl.\ Phys.\ B {\bf 316} (1989) 391.
\bibitem{sgrev1}T.~Battefeld and S.~Watson,
  {\it String gas cosmology},
  Rev.\ Mod.\ Phys.\  {\bf 78} (2006) 435
  [\arXiv{hep-th/0510022}].
\bibitem{sgrev2} R.~H.~Brandenberger,
  {\it String gas cosmology},
  \arXiv{0808.0746 [hep-th]}; {\it String gas cosmology: progress and problems},
  Class.\ Quant.\ Grav.\  {\bf 28} (2011) 204005
  \arXiv{1105.3247 [hep-th]}.
\bibitem{qbackreact} O. Evnin, {\it Quantum  backreaction in string theory}, Fortschr. Phys. {\bf 60} (2012) 998-1004 [\arXiv{arXiv:1201.6606}].
  \bibitem{intcom}R.~Danos, A.~R.~Frey and A.~Mazumdar,
  {\it Interaction rates in string gas cosmology}, Phys.\ Rev.\ D {\bf 70} (2004) 106010
  [\arXiv{hep-th/0409162}].
 \bibitem{Friedan_IR} D. Friedan, {\it A tentative theory of large distance physics}, \\ JHEP 0310 (2003) 063 [\arXiv{hep-th/0204131}].
\bibitem{polch}J.~Polchinski, {\it String theory} (CUP, 1998) v.~1. 
\bibitem{polchinski-fischler-susskind}J.~Polchinski,
  {\it Factorization of bosonic string amplitudes},
  Nucl.\ Phys.\ B {\bf 307}, 61 (1988).
\bibitem{pt}V.~Periwal and O.~Tafjord,
  {\it D-brane recoil},
  Phys.\ Rev.\ D {\bf 54} (1996) 3690
  [\arXiv{hep-th/9603156}].
\bibitem{hk} S.~Hirano and Y.~Kazama,
  {\it Scattering of closed string states from a quantized D particle},
  Nucl.\ Phys.\ B {\bf 499} (1997) 495
  [\arXiv{hep-th/9612064}].
\bibitem{worldline}O.~Evnin,
  {\it Worldline techniques for string theory solitons},
  \arXiv{hep-th/0507180}.
\bibitem{ct} C.~G.~Callan and L.~Thorlacius,
  {\it Sigma Models And String Theory},
  in proceedings of {\it Providence 1988: Particles, strings and supernovae}, v.~2, pp.~795-878, http://www.damtp.cam.ac.uk/user/tong/string/sigma.pdf
  
  \bibitem{Tseytlin_rev} A. Tseytlin, {\it Sigma model approach to string theory},\\
Int. J. of Mod. Phys. vol.~4 no.~6 (1989) 1257. 
  
\bibitem{isham}C.~J.~Isham,
  {\it Canonical quantum gravity and the problem of time},
  in proceedings of {\it Salamanca 1992}, pp. 157-287 [\arXiv{gr-qc/9210011}].
\bibitem{halliwell}J.~J.~Halliwell,
  {\it Introductory lectures on quantum cosmology},
  in Jerusalem 1989, Proceedings, Quantum cosmology and baby universes, p. 159-243, 
  [\arXiv{0909.2566 [gr-qc]}].
\bibitem{HH}J.~B.~Hartle and S.~W.~Hawking,
  {\it Wave function of the universe},\\
  Phys.\ Rev.\ D {\bf 28} (1983) 2960.

\bibitem{marolf}D.~Marolf,
  {\it Path integrals and instantons in quantum gravity: Minisuperspace models},
  Phys.\ Rev.\ D {\bf 53} (1996) 6979
  [\arXiv{gr-qc/9602019}].
\bibitem{hllwll}
  J.~J.~Halliwell,
  {\it Derivation of the Wheeler-De Witt equation from a path integral for minisuperspace models},
  Phys.\ Rev.\ D {\bf 38} (1988) 2468.
  \bibitem{Misner} C. W. Misner, in {\it Magic without magic: John Archibald Wheeler} (Freeman, San Francisco, 1972), pp. 441-473; {\it A minisuperspace example: the Gowdy $T^3$ cosmology}, 
  Phys. Rev. D {\bf 8} (1973) 3271. 
  
\bibitem{friedan} D. Friedan, {\it Nonlinear models in $2+\epsilon$ dimensions}, Ph.D. thesis, University of California at Berkeley (August 1980),  Ann. Phys. (NY) {\bf 163} (1985) 318.
  
\bibitem{CMNP} A. Cohen, G. Moore, P. Nelson, and J. Polchinski,\\ 
{\it Semi-off-shell string amplitudes}, Nucl.\ Phys.\ B {\bf 281} (1987) 127;\\ 
{\it An off-shell propagator for string theory},
Nucl.\ Phys.\ B {\bf 267} (1986) 143. 

\bibitem{linear}H.~Fukutaka,
  {\it Path integral measure of linearized gravity in curved space-time},\\
  \arXiv{hep-th/9203049}.
\bibitem{cd} S.~M.~Christensen and M.~J.~Duff,
  {\it Quantizing gravity with a cosmological constant},
  Nucl.\ Phys.\ B {\bf 170} (1980) 480.
\bibitem{hamber} H.~W.~Hamber, {\it Quantum Gravitation} (Springer, 2008), p. 145.
\bibitem{Lichnerowicz} A.~Lichnerowicz, {\it Propagateurs et commutateurs en relativit\'e g\'en\'erale},\\ Pub. Mat. IHES {\bf 10} (1961) 5.
\bibitem{Ricciflow} B.~Chow, P.~Lu, L.~Ni, {\it Hamilton's Ricci flow} (AMS, 2006), p. 109.
\bibitem{TseytlinRG}A.~A.~Tseytlin,
  {\it On sigma model RG flow, `central charge' action and Perelman's entropy},
  Phys.\ Rev.\ D {\bf 75} (2007) 064024
  [\arXiv{hep-th/0612296}].
\bibitem{JMW} S.~Jain, G.~Mandal and S.~R.~Wadia,
  {\it Virasoro conditions, vertex operators and string dynamics in curved space},
  Phys.\ Rev.\ D {\bf 35} (1987) 778;  {\it Perturbatively renormalized vertex operator, highest weight representations of Virasoro algebra and string dynamics in curved space},
  Phys.\ Rev.\ D {\bf 35} (1987) 3116.
\bibitem{kasner1}M.~T.~Mueller, {\it Rolling radii and a time-dependent dilaton},\\
  Nucl.\ Phys.\ B {\bf 337} (1990) 37.
  \bibitem{kasner3} A. Tseytlin, {\it String cosmology and dilaton},  \arXiv{hep-th/9206067}.
\bibitem{kasner2}D.~L\"ust,
  {\it Cosmological string backgrounds},
  \arXiv{hep-th/9303175}.
\end{thebibliography}
\end{document}